
\input amstex
\magnification 1200
\documentstyle{amsppt}
\NoBlackBoxes
\NoRunningHeads
\def\g{\frak g}
\def\sltwo{\frak s\frak l _2 }
\def\Z{\Bbb Z}
\def\C{\Bbb C}
\def\R{\Bbb R}
\def\d{\partial}
\def\i{\text{i}}
\def\ghat{\hat\frak g}
\def\gtwisted{\hat{\frak g}_{\gamma}}
\def\gtilde{\tilde{\frak g}_{\gamma}}
\def\Tr{\text{\rm Tr}}

\topmatter
\title Representations of affine Lie algebras, parabolic differential
equations  and Lam\'e functions
  \endtitle
\author {\rm {\bf Pavel I. Etingof, Alexander A. Kirillov, Jr.} \linebreak
	\vskip .1in
   Department of Mathematics\linebreak
   Yale University\linebreak
   New Haven, CT 06520, USA\linebreak
   e-mail: etingof\@math.yale.edu,
			kirillov\@math.yale.edu}
\endauthor

\abstract

We consider correlation functions for the Wess-Zumino-Witten model
on the torus with the insertion of a Cartan element; mathematically
this means that we consider the function of the form $F=\Tr (\Phi_1
(z_1)\ldots \Phi_n (z_n)q^{-\d}e^{h})$ where $\Phi_i$ are intertwiners
between Verma modules and evaluation modules over an affine Lie
algebra $\ghat$, $\d$ is the grading operator in a Verma module and
$h$ is in the Cartan subalgebra of $\g$. We derive a system of
differential equations satisfied by such a function. In particular,
the calculation of $q\frac{\d} {\d q} F$ yields a parabolic second
order PDE closely related to the heat equation on the compact Lie
group corresponding to $\g$ (cf. \cite{Ber}). We consider in detail
the case $n=1$, $\g = \sltwo$. In this case we get the following
differential equation ($q=e^{\pi \i \tau}$): $ \left( -2\pi\i
(K+2)\frac{\d}{\d\tau}+\frac{\d^2}{\d x^2}\right) F = (m(m+1)
\wp(x+\frac{\tau}{2}) +c)F$, which for $K=-2$ (critical level) becomes
Lam\'e equation. For the case $m\in\Z$ we derive integral formulas for
$F$ and find their asymptotics as $K\to -2$, thus recovering classical
Lam\'e functions.
\endabstract

\date October 4, 1993 \enddate

\endtopmatter

\document

\heading {\bf Introduction.}\endheading

We start with consideration of the Wess-Zumino-Witten model of
conformal field theory on a torus. This is, we consider an affine Lie
algebra $\ghat$ corresponding to some simple finite-dimensional Lie
algebra $\g$. For technical reasons, it is more convenient to work
with a twisted realization of $\ghat$. Next, we consider Verma modules
$M_{\lambda, k}$ over
$\ghat$. If $V$ is a representation of the finite-dimensional algebra
$\g$ then by definition a vertex operator $\Phi(z)\colon M_{\lambda,
k}\to M_{\nu,k}\otimes V$ is an operator valued formal Laurent series
in $z$ satisfying the following commutation relations with the
elements of $\ghat$:

$$\Phi(z) a\otimes t^m = \left( (a\otimes t^m) \otimes 1 +z^m 1\otimes
a\right) \Phi(z).$$


Let $M_{\lambda_i,k},
i=0\ldots n$ be a collection of Verma modules such that
$\lambda_0=\lambda_n$, and $\Phi^i(z_i):M_{\lambda_i,k}\to
M_{\lambda_{i-1}, k}\otimes V_i$ be vertex operators. Then we can consider
the following ``correlation function on the torus'':

$$\Cal F (z_1\ldots z_n,q,h)= \Tr|_{M_{\lambda_0, k}}\left(
\Phi^1(z_1)\ldots \Phi^n(z_n) q^{-\d} e^h\right) $$
where $\d$ is the grading operator in Verma modules\footnote{We use
the symbol $\d$ for the grading operator in twisted realization,
reserving the standard notation $d$ for the untwisted grading
operator, see Section 1.} and $h\in \frak h _\R$. This
function takes values in the module $V=V_1\otimes\ldots \otimes V_n$
and it is the main object of our study.

Our first goal is to derive differential equations for $\Cal F$.  We
compute
$\frac{\d}{\d z_i}\Cal F$
using the same technique as for the usual
Knizhnik-Zamolodchikov equations (see \cite{TK}, \cite{FR}). However,
this system of equations (Theorem 3.1) is not closed: it has the form

$$z_i\frac{\d}{\d z_i} \Cal F = A_i(z_1\ldots z_n)\Phi +
	\sum \pi_i(x_l)\frac{\d}{\d x_l} \Phi$$
where $A_i$ are some operators in $V$ and the sum is taken over
an orthonormal basis $x_l$ in $\frak h$. Since we do not have any
information about $\frac{\d}{\d x_l} \Cal F$, this system does not allow
us  to determine $\Cal F$. This system of equations in another form
appeared first in the papers of Bernard (\cite{Ber}).

However, we can get additional information studying dependence on the
modular parameter $q$. This dependence takes a simpler form if we
renormalize $\Cal F$, considering $\Cal F/\Cal F_0$, where $\Cal F_0 =
\Tr|_{M_{0,k}} (q^{-\d}e^{h})$. For this function we have the following
equation (Theorem 4.1):

$$(q\frac{\d}{\d q} - \frac{1}{2(k+1)}\Delta_{\frak
h})\frac{\Cal F}{\Cal F_0}= \frac{1}{k+1}B\frac{\Cal F}{\Cal F_0},$$
where $B$ is some operator in $V$ depending on $z_i,q$.

This equation is especially interesting for $n=0$ and $n=1$,
since in these cases there is no dependence on $z$. For $n=0$ we
obtain that $\Cal F/\Cal F_0$ satisfies the heat equation on the maximal
torus $T\subset G$. This result was first obtained by Bernard;
however, he worked in the untwisted setting which gave him heat equation
on the group rather than on the torus.

We concentrate our attention on the case $n=1, \g =\sltwo$ and take
$V$ to be $2m+1$-dimensional  irreducible module over $\sltwo$.
In
this case the intertwiner $\Phi(z)\colon M_{\lambda,k}\to
M_{\lambda,k}\otimes V$ exists  and
is unique up to a constant for generic values of
$k,\lambda$. Therefore we can apply the results
of the previous sections. It will be convenient to identify
$\frak h_\R\simeq \R$ and consider the following function
of one real variable $x$:

$$F(x,q)=\frac{\Tr|_{M_{\lambda, k}} \left( \Phi(z)q^{-\d}e^{-\pi\i x
h}\right) }
{\Tr|_{M_{0,k}} \left( q^{-\d}e^{-\pi \i x h}\right)} ,$$
where $h=\left(\matrix 1&0\\ 0&-1 \endmatrix\right)\in\sltwo$.
Then the results of Section 4 show that thus defined $F$ satisfies the
following partial differential equation (eq. (5.1)):

$$\left(-4\pi \i(k+1)\frac{\d}{\d \tau} +\frac{\d ^2}{\d
x^2}\right)F=m(m+1)\left(\wp\left(\frac{\tau}{2} +x\right)+c\right)
F,$$
where $q=e^{\pi \i\tau}$.

This equation has many nice properties and is worth studying in
itself; it is a non-stationary Schr\"odinger equation. Equations of
this type were considered and treated in \cite{Kri} and references
therein. What is particularly interesting is that in the limit $k\to
-1$ this equation becomes the Lam\'e equation with precisely the same
expression of the coefficient to $\wp$ as in the classical form (i.e.,
$m(m+1)$, cf. \cite{WW}). Note
that the case $k=-1$ in the twisted setting corresponds to $K=-2$ in
the untwisted one, i.e. to the critical level. Therefore we should
expect that
the asymptotics of the correlation functions on the torus at the
critical level limit are closely related to the solutions of Lam\'e
equation.

We make this relation more explicit.
First we derive integral formulas for the function $F$ defined above
using the  Wakimoto realization of Verma modules over $\widehat{\sltwo}$.
The answer is given by  Theorem 5.1 and has the following form:

$$F(x,q)=\int\limits_{\Delta}
l(\zeta|\tau)^{1/\kappa}\phi(x,\zeta|\tau) d\zeta_1\ldots d\zeta_m $$
where both $l(\zeta)$ and $\phi(x,\zeta)$ are expressed in terms of
theta functions and exponents and $\kappa=2(k+1)$.

Now we can find the asymptotics of these functions as $\kappa\to 0$.
To do it we need the saddle-point method which says that under certain
assumptions this integral has the following asymptotics:

$$F(x,q)\sim C(\tau) \kappa^{m/2} l(\zeta_0|\tau)^{1/\kappa}
\phi(x,\zeta_0|\tau), $$
where $\zeta_0$ is a critical point for the function $l(\zeta)$.
Substituting this asymptotics in  equation (5.1) we finally see
that the function

$$f(x)=\phi(x,\zeta_0)$$
is a solution of  the Lam\'e equation. Note that we cannot find the
critical point $\zeta_0$ explicitly; we can only write a system of
equations defining it; this system of equations will also depend on
the highest weight $\lambda$ of the Verma module.

It turns out that in this way we can get a generic eigenfunction of
the Lam\'e operator; in particular, we can get all the doubly periodic
eigenfunctions
which are obtained for integer $\lambda$. For example, for $m=1$ we
show how to get the classical solutions $f(x)= \text{dn } 2\bold K x,
\text{sn }2\bold K x, \text{cn }2\bold K x,$ which correspond to
even (the first solution) and odd (two latter solutions) values of
$\lambda$.

However, it is not very easy to check the applicability of the
saddle-point method. For the case $m=1$, we can check it in one
special case ($q\in\R, \lambda$ is small) and then use analytic continuation
arguments to show that in fact every critical point gives rise to an
eigenfunction of Lam\'e equation. In the general case one could repeat
similar arguments to prove that again any critical point gives a
solution of Lam\'e eqution; however, we do not follow this way;
instead, we say that it can be checked by direct calculation which is
in fact carried out in the book by Whittaker and Watson \cite{WW}.

\remark{Remark}
The asymptotics of solutions of the KZ equations at the critical level
were  computed by means of the saddle-point method by Reshetikhin and
Varchenko\cite{RV}. They found that the leading term of these asymptotics
is  a solution of the Bethe equations well known in physics.
\endremark

\heading{\bf Acknowledgements}\endheading

We  are deeply grateful to our advisor professor Igor
Frenkel who suggested this problem to us . Without his encouragement
and inspiring discussions with him this paper would never be written.
We would also like to thank professors I. Cherednik, B. Feigin, H. Garland,
M. Jimbo, M. Kashiwara, T. Miwa, F. Malikov and A. Varchenko for helpful
discussions.

The work of P.E. was supported by the Alfred P. Sloan graduate
dissertation fellowship.

\heading
{\bf 1. A twisted realization of affine Lie algebras.} \endheading

Let $\frak g$ be a finite dimensional simple Lie algebra over $\Bbb C$
of rank $r$.  Denote by $<,>$ the standard invariant form on $\frak g$
with respect to which the longest root has length $\sqrt{2}$.

Let $\frak h$ denote a Cartan subalgebra of $\frak g$. The form $<,>$
defines a natural identification ${\frak h^*}\to{\frak h}$:
$\lambda\mapsto h_{\lambda}$ for $\lambda\in{\frak h}^*$. We will use
the notation $<,>$ for the inner product in both $\frak h$ and $\frak
h^*$.

Let $\Delta^+$ be the set of positive roots of $\frak g$.  For
$\alpha\in\Delta^+$, let $e_{\alpha}\in\g_{\alpha}$,
$f_{\alpha}\in\g_{-\alpha}$ be such that $<e_\alpha,f_\alpha>=1$. Then
$[e_{\alpha},f_{\alpha}]=h_{\alpha}$.  Also, let $x_j$, $1\le j\le r$,
be an orthonormal basis of $\frak h$ with respect to the standard
invariant form. Then the elements $e_{\alpha}, f_{\alpha}, x_i$ form a
basis of $\g$.

Let $g$ be the dual Coxeter number of $\frak g$. Let
$\rho=\frac{1}{2}\sum_{\alpha\in\Delta^+}\alpha$.  Then for
$\alpha\in\Delta^+, <~\rho,\alpha>=|\alpha|$ is the number of summands
in the decomposition of $\alpha$ in the sum of simple positive roots.

Let $\gamma=\text{Ad }e^{\frac{2\pi \text{i}h_{\rho}}{g}}$ be an
inner  automorphism of $\frak g$.
\footnote{Note that throughout the
paper we denote the complex number $\text{i}=\sqrt{-1}$ by a roman
``i'', to distinguish it from the subscript $i$, which is italic.}
This automorphism is of order $g$.

The action of $\gamma$ on root vectors is as follows:
$\gamma(e_{\alpha})=\varepsilon^{|\alpha|}e_{\alpha}$ ,
$\gamma(f_{\alpha})=\varepsilon^{-|\alpha|}f_{\alpha}$ , where
$\varepsilon=e^{2\pi \text{i}/g}$ is a primitive $g$-th root of
unity.  Note that since $0<|\alpha|\le g-1$, $\gamma(a)=a$
if and only if $a\in \frak h$.

Let $\hat{\frak g}={\frak g}\otimes \Bbb C[t,t^{-1}]\oplus \Bbb Cc$ be
the affine Lie algebra associated with $\frak g$. The commutation
relations in this algebra are $$ [a(t)+\lambda c,b(t)+\mu
c]=[a(t),b(t)]+\frac{1}{2\pi \text{i}}
\oint_{|t|=1}<a^{\prime}(t)b(t)>t^{-1}dt\cdot c\tag 1.1
$$ for any two $\frak g$-valued Laurent polynomials $a(t)$, $b(t)$,
and complex numbers $\lambda$, $\mu$.

Define the twisted affine algebra $\hat{\frak g}_{\gamma}$ as a
subalgebra of $\hat{\frak g}$
consisting of all expressions $a(t)+\lambda c$ with the property
$a(\varepsilon t)=\gamma(a(t))$. The basis in this subalgebra is given
by elements $e_{\alpha}\otimes t^{|\alpha|+mg}$, $f_{\alpha}\otimes
t^{-|\alpha|+mg} $, $x_i\otimes t^{mg}$, $c$, for $m\in \Bbb Z$,
$\alpha\in\Delta^+$. For brevity we  write $x[n]$ for $x\otimes t^n$
and $x$ for $x\otimes t^0$.

The representation theory of $\hat{\frak g}_{\gamma}$ is
quite parallel to that in the untwisted case.  First of all,
define the polarization of $\hat{\frak g}_{\gamma}$: $\hat{\frak
g}_{\gamma}=\hat{\frak g}_{\gamma}^+\oplus\hat{\frak g}_{\gamma}^-
\oplus {\frak h}\oplus \Bbb Cc$. Here $\hat{\frak g}_{\gamma}^+$ is the set of
polynomials $a(t)$ vanishing at $0$, and $\hat{\frak g}_{\gamma}^-$ is
the set of polynomials $a(t)$ vanishing at infinity.

Next, define Verma modules over $\hat{\frak g}_{\gamma}$. This is done
exactly in the same way as for the untwisted affine algebra.  Let
$\lambda\in \frak h^*$ be a weight, and let $k$ be a complex number.
Define $X_{\lambda,k}$ to be a one dimensional module over $\hat{\frak
g}_{\gamma}^+\oplus {\frak h}\oplus \Bbb Cc$ spanned by a vector $v$
such that $\hat{\frak g}_{\gamma}^+$ annihilates $v$, and $cv=kv$,
$hv=\lambda(h)v$, $h\in \frak h$. Define the Verma module $$
M_{\lambda,k}=\text{Ind}^{\hat{\frak g}_{\gamma}}_{\hat{\frak
g}_{\gamma}^+
\oplus {\frak h}\oplus \Bbb Cc}X_{\lambda,k}.\tag 1.2
$$

Now define evaluation representations.  Let $V$ be a module over
$\frak g$; we always assume that $V$ has a weight decomposition such
that weight subspaces are finite-dimensional.  Define the operator
$C\in\text{End}(V)$ by $C=e^{\frac{2\pi\text{i} h_\rho}{g}}$; then
$Caw=\gamma(a)Cw$ for any $w\in V$, $a\in\frak g$, and $Cw_0=w_0$, if
$w_0$ is from the zero-weight subspace $V(0)$.

Let $V(z)$ denote the space of $V$-valued Laurent polynomials in $z$,
and let $V_C(z)$ be the space of those polynomials which satisfy the
equivariance condition $w(\varepsilon z)=Cw(z)$.

The natural (pointwise) action of $\hat{\frak g}$ on $V(z)$ restricts
to an action of $\hat{\frak g}_{\gamma}$ on $V_C(z)$.

Note that the twisted affine algebra is isomorphic to untwisted one.
More precisely, we have the following lemma.

\proclaim {Lemma 1.1 \rm (see \cite{PS},p.36)} The Lie algebras
$\hat{\frak g}_{\gamma}$ and $\hat{\frak g}$ are isomorphic. Under
this isomorphism, the Verma module $M_{\lambda,k}$ over $\gtwisted$ is
identified with the Verma module $\Cal M_{\lambda+k\rho, gk}$ over
$\ghat$, and evaluation representation $V_C(z)$ over $\gtwisted$ is
identified with the evaluation representation $V(z)$ over $\ghat$.
\endproclaim

Therefore, all the results about the representations of $\gtilde$ can
be as well obtained from the representations theory of $\tilde\g$.
However, use of the twisted algebra is technically more convenient to
us.

We can also extend the Lie algebra $\gtwisted$ adding to it the degree
operator $\d$ which commutes with elements of $\gtwisted$ by $[\d,
a(t)]=ta'(t), [\d,c]=0$. We denote this extended Lie algebra by
$\gtilde$. Note that under the isomorphism of Lemma 1.1, $\d\mapsto
gd+h_\rho$, where $d$ is the standard grading operator for $\hat\g$.

Since every Verma module has a natural $\Z$-gradation, we can uniquely
extend the action of $\gtwisted$ in $M_{\lambda,k}$ to the action of
$\gtilde$ as soon as we define the action of $\d$ on the highest
weight vector. Let us define it by
$\d v_{\lambda,k} =-\frac{<\lambda,\lambda>}{2(k+1)}v_{\lambda,k}$.
Again, this agrees with the untwisted case: if we identify
$M_{\lambda, k}$ with the Verma module $\Cal M _{\Lambda, K}$ over
$\hat\g$, where $\Lambda=\lambda+k\rho, K=gk$, then $\d$ is identified
with $gd+h_\rho-\frac{k^2<\rho,\rho>}{2(k+1)}$, where the action of
$d$ on Verma module $\Cal M_{\lambda, K}$ is defined so that on the
highest level, $d=-\frac{<\Lambda, \Lambda+2\rho>}{2(K+g)}$.
This gives us the following twisted analogue of
Sugawara construction for $\d$.
\proclaim{Proposition 1.2 \rm (see \cite{E})} In every Verma module over
$\gtwisted$,
$$
\d=-\frac{1}{k+1}\sum_{m\in\Bbb Z}\left(\sum_{\alpha\in\Delta^+}
:e_{\alpha}[|\alpha|+mg]f_{\alpha}[-|\alpha|-mg]:
+\frac{1}{2}\sum_{j=1}^{r}:x_j[mg]x_j[-mg]:\right)
,\tag 1.3 $$ where :: is the standard normal ordering: $$\gather
:e_{\alpha}[n]f_{\alpha}[-n]:=\cases
e_{\alpha}[n]f_{\alpha}[-n] & n<0\\
f_{\alpha}[-n]e_{\alpha}[n] & n>0\endcases
\\
:h[n]h[-n]:=\cases h[n]h[-n]&n\le
0\\ h[-n]h[n] &n>0 \endcases,\ h\in{\frak h}.
\tag 1.4
\endgather
$$
\endproclaim

We can also define the action of $\d$ in the evaluation representation
$V_C(z)$ by $\d=z\frac{d}{dz}$. Thus $V_C(z)$ becomes a
$\gtilde$-module. Again, under the isomorphism of the Lemma 1.1, $\d$
is identified with the operator $gd+h_\rho$ in the evaluation
representation $V(z)$ over $\hat\g$, where we define the action of $d$
in $V(z)$ by $d=z\frac{d}{dz}$. \footnote{ Note, however, that usually
in the conformal field theory the action of $d$ on the evaluation
reperesentation is defined with some shift, namely: $d=z\frac{d}{dz} +
\frac{C}{2(K+g)}$, where $C$ is the Casimir operator. We will note use
this convention here.}

Let us introduce the twisted version of currents.  Set $$
\gather
J_{e_{\alpha}}(z)=\sum_{m\in \Bbb Z}e_{\alpha}
[|\alpha|+mg]\cdot z^{-|\alpha|-mg-1},\\
J_{f_{\alpha}}(z)=\sum_{m\in \Bbb Z}f_{\alpha}
[-|\alpha|+mg]\cdot z^{|\alpha|-mg-1},\\
J_h(z)=\sum_{m\in \Bbb
Z}h[mg]\cdot z^{-mg-1},\ h\in{\frak h}.\tag 1.5\endgather $$
Thus by linearity we have defined $J_a(z)$ for any $a\in {\frak g}$.

Define the polarization of currents: $$
\gather
J^+_{e_{\alpha}}(z)=\sum_{m<0}e_{\alpha}[|\alpha|+mg]\cdot
z^{- |\alpha|-mg-1},\\
J^+_{f_{\alpha}}(z)=\sum_{m\le
0}f_{\alpha}[-|\alpha|+mg]\cdot z^{|\alpha|-mg-1},\\
J^+_h(z)=\frac{1}{2}h\cdot z^{-1}+\sum_{m<0}h
[mg]\cdot z^{- mg-1},\ h\in{\frak h}.\tag 1.6\endgather $$
This defines $J_a^+(z)$ for all $a\in\frak g$. Now set
$$J_a^-(z)=J_a^+(z)-J_a(z).\tag 1.7 $$

Note that this polarization is not quite the same as the standard
polarization of currents for the untwisted $\hat{\frak g}$,
i.e. the isomorphism between $\ghat$ and $\gtwisted$ does not match up
these two polarizations.

One can easily write commutation relations between currents; however,
we won't need them. One thing we will need is the commutation
relations with $\d$. Namely, one can easily see that $$
J^\pm_a(z)q^{-\d}=q^{-1} q^{-\d}J^\pm (q^{-1}z)\tag 1.8$$

\vskip 0.1in
\heading{\bf 2. Twisted intertwiners and correlation functions on a torus
} \endheading

We will be interested in $\tilde{\frak g}_{\gamma}$ intertwining
operators $\Phi(z):M_{\lambda,k}\to M_{\nu,k}\hat\otimes
z^{\Delta}V_C(z)$, where the highest weight of $V$ is $\mu$,
$\hat\otimes$ denotes the completed tensor product, and $\Delta$ is a
complex number.

Let $z_0$ be a nonzero complex number. Evaluation of the operator
$\Phi(z)$ at the point $z_0$ yields an operator
$\Phi(z_0):M_{\lambda,k}\to
\hat M_{\nu,k}\otimes V$, where $\hat M$ denotes the completion of $M$
with respect to the grading.

 From now on the notation $\Phi(z)$ will mean the operator $\Phi$
evaluated at the point $z\in\Bbb C^*$. This will give us an
opportunity to regard the operator $\Phi(z)$ as an analytic function
of $z$. This analytic function will be multivalued:
$\Phi(z)=z^{\Delta}\Phi^0(z)$, where $\Phi^0$ is a single-valued
function in $\Bbb C^*$, and
$\Delta=\frac{<\nu,\nu>-<\lambda,\lambda>}{2(k+1)}$.

Let $u$ belong to the restricted dual module $V^*$. Introduce the
notation
\linebreak $\Phi_u(z)=(1\otimes u)(\Phi(z))$. $\Phi_u(z)$ is an
operator: $M_{\lambda,k}\to
\hat M_{\nu,k}$.

The intertwining property for $\Phi(z)$ can be written in the form $$
[a[m],\Phi_u(z)]=z^m\Phi_{au}(z).\tag 2.1 $$

It is convenient to write the intertwining relation in terms of
currents.

\proclaim{Lemma 2.2}
$$
\gather
[J_h^{\pm}(\zeta),\Phi_u(z)]=\zeta^{-1}x^0(\zeta/z)
\Phi_{h u}(z);\\
[J_{e_{\alpha}}^{\pm}(\zeta),\Phi_u(z)]=\zeta^{-1}x^\alpha(\zeta/z)
\Phi_{e_{\alpha}u}(z),\ \alpha\in\Delta_+;\\
[J_{f_{\alpha}}^{\pm}(\zeta),\Phi_u(z)]=\zeta^{-1}x^{-\alpha}(\zeta/z)
\Phi_{f_{\alpha}u}(z),\ \alpha\in\Delta_+,\tag 2.2\endgather
$$ where $$\gather x^0(t)=\frac{1}{2}\frac{1+t^g}{1-t^g},\\
x^\alpha(t)=t^{-|\alpha|}\frac{t^g}{1-t^g},\\
x^{-\alpha}(t)=t^{|\alpha|}\frac{1}{1-t^g}\tag 2.3\endgather $$

\endproclaim

The identities marked with $+$ make sense if $|z|>|\zeta|$, and those
marked with $-$ make sense if $|z|<|\zeta|$.

Now we are ready to write down the twisted version of the operator
Knizhnik- Zamolodchikov (KZ) equations.

\proclaim{Theorem 2.3 \rm(see\cite{E})} The operator function $\Phi_u(z)$
satisfies the
differential equation $$
\gather
(k+1)\frac{d}{dz}\Phi_u(z)=\sum_{\alpha\in\Delta^+}
(J_{e_{\alpha}}^+(z)\Phi_{f_{\alpha}u}(z)-\Phi_{f_{\alpha}u}(z)
J_{e_{\alpha}}^-(z))+\\
\sum_{\alpha\in\Delta^+}
(J_{f_{\alpha}}^+(z)\Phi_{e_{\alpha}u}(z)-\Phi_{e_{\alpha}u}(z)
J_{f_{\alpha}}^-(z))+
\sum_{j=1}^{r}
(J_{x_j}^+(z)\Phi_{x_ju}(z)-\Phi_{x_ju}(z) J_{x_j}^-(z)).\tag
2.4\endgather $$
\endproclaim

In the paper \cite{E} this proposition is proved in the case when $V$
is a highest-weight module. However, one can check that the result is
valid for any module with the weight decomposition.

\remark{Remark} Note that since we have the isomorphism of Lemma 1.1,
we can identify the intertwiners for $\gtilde$ with those for
$\tilde\g$. Namely, if $\hat\Phi(\zeta)\colon M_{\Lambda,K}\to M_{N,
K}\otimes V_{\mu}$ is an intertwiner for $\tilde\g$, then the results
of Section 1 show that

$$\Phi(z)=z^{-h_\rho}\hat\Phi(z^g)\tag 2.5$$ will be an intertwiner
for $\gtilde$.\endremark
\vskip 0.1in

The main object of our study will be the following twisted version of
correlation functions on a torus. Let $V_1,...,V_n$ be representations
of $\frak g$, and let $\Phi^i(z_i): M_{\lambda_i,k}\to \hat
M_{\lambda_{i-1},k}\otimes V_i$, $1\le i\le n$ be intertwining
operators for $\gtilde$(sometimes for brevity we write $\Phi^i$
instead of $\Phi^i_{u_i}(z_i)$). If $\lambda_0=\lambda_n=\lambda$,
then we can define the following correlation function on a torus:
$$\Cal F_{u_1,\ldots,u_n}(z_1,\ldots, z_n;h) =
\Tr|_{M_{\lambda, k}}(\Phi^1_{u_1}(z_1)\ldots
\Phi^n_{u_n}(z_n)q^{-\d}e^h),\tag 2.6$$
where $q\in \C^\times, |q|<1$, $h\in\frak h_\R$ is an element from the
compact form of $\frak h$ (i.e., $h(\alpha)\in \text{i} \R$ for all
$\alpha \in\Delta$). This trace is a formal power series in $z_1\ldots
z_n$.  However, it can be shown that it converges to an analytic
function of $z_i$ in the region $|z_1|>\ldots >|z_n|>|qz_1|$.
Therefore, we will consider $\Cal F$ as a function of
$q,z_1,\ldots,z_n,h$ with values in $V_1\otimes\ldots \otimes V_n$ or,
equivalently, as a function of $q,z_i$ with values in $V_1\otimes
\ldots \otimes V_n\otimes C^{\infty}(\frak h_\R)$.  Our first goal is finding
differential equations for this function.

\vskip 0.1in
\heading{\bf 3. Differential equations for correlation functions.}\endheading

In this section we deduce the differential equations for the
correlation function on the torus $\Cal F$ defined by (2.6). The basic
tool will be the application of operator KZ equation (2.3), which
implies:
$$\gather (k+1)z_i \frac{\d}{\d z_i}\Cal F_{u_1,\ldots ,
u_n}(z_1,\ldots, z_n)=\\
\sum\limits_{\alpha\in\Delta^+}z_i \Tr (\Phi^1\ldots
(J^+_{e_\alpha}(z_i)\Phi^i_{f_\alpha u_i}(z_i) -\Phi^i_{f_\alpha
u_i}(z_i)J^-_{e_\alpha} (z_i))\ldots \Phi^n q^{-\d}e^h) \\
+\sum\limits_{\alpha\in\Delta^+}z_i \Tr (\Phi^1\ldots
(J^+_{f_\alpha}(z_i)\Phi^i_{e_\alpha u_i}(z_i) -\Phi^i_{e_\alpha
u_i}(z_i)J^-_{f_\alpha} (z_i))\ldots \Phi^n q^{-\d}e^h) \\
+\sum\limits_{l=1}^r z_i\Tr (\Phi^1\ldots (J^+_{x_l}(z_i)\Phi^i_{x_l
u_i}(z_i) -\Phi^i_{x_l u_i}(z_i)J^-_{x_l} (z_i))\ldots \Phi^n q^{-\d}e^h).
\tag 3.1\endgather$$

Let us consider the first summand in (3.1). Using commutation
relations of currents with intertwiners (2.2), we can move
$J^+_{e_\alpha}(z_i)$ to the utmost left, and $J^-_{e_\alpha}(z_i)$ to
the utmost right, which gives us the following expression for the
first term in (3.1):
$$\split -\sum\limits_{\alpha\in\Delta^+}\biggl(
&\sum\limits_{j\ne i} x^\alpha(z_i/z_j) \Tr (\Phi^1\ldots
\Phi^j_{e_\alpha u_j}(z_j)\ldots \Phi^i_{f_\alpha u_i}(z_i)\ldots
\Phi^n q^{-\d}e^h)\biggr.\\
&-z_i\Tr (J^+_{e_\alpha}(z_i)\Phi^1\ldots \Phi^i_{f_\alpha u_i}(z_i)\ldots
\Phi^n q^{-\d}e^h)\\
&\biggl.+z_i\Tr (\Phi^1\ldots \Phi^i_{f_\alpha u_i}(z_i)\ldots
\Phi^nJ^-_{e_\alpha}(z_i)q^{-\d}e^h)\biggr).\endsplit\tag 3.2$$

Now, using the cyclic property of trace and the relation $q^{-\d}e^h
J^\pm_{e_\alpha}(z)=qe^{h(\alpha)}J^\pm_{e_\alpha}(qz)q^{-\d}e^h$, we
can rewrite (3.2) as follows:

$$\split -\sum\limits_{\alpha\in\Delta^+}
\biggl(&\sum\limits_{j\ne i}x^\alpha(z_i/z_j)
\pi_i\otimes \pi_j (f_\alpha\otimes e_\alpha) \biggr.\Phi\\
&-e^{h(\alpha)} qz_i\Tr (\Phi^1\ldots \Phi^i_{f_\alpha u_i}(z_i)\ldots
\Phi^nJ^+_{e_\alpha}(qz_i)q^{-\d}e^h)\\
&+e^{-h(\alpha)}q^{-1}z_i\Tr (J^-_{e_\alpha}(q^{-1}z_i)\Phi^1
\ldots \Phi^i_{f_\alpha u_i}(z_i)\ldots
\biggl.\Phi^nq^{-\d}e^h)\biggr)\endsplit$$

Repeating this procedure and taking into account that
$\operatornamewithlimits{lim}\limits_{m\to +\infty} q^{\pm m}z
J^\pm_{e_\alpha} (q^{\pm m}z) =0$, we finally get the following
expression for the first summand in (3.1):

$$ -\sum\limits_{\alpha\in\Delta^+}
{\sum\limits_{m\in\Z,j}}'x^\alpha(q^mz_i/z_j)e^{mh(\alpha)}
\pi_i\otimes \pi_j(f_\alpha\otimes
e_\alpha)\Phi,\tag 3.3$$
where $\sum'$ means that the sum is taken
over all $m\in\Z, j=1\ldots n$ except $m=0, j=i$. (We use the
convention $\pi_i \otimes \pi_i(a\otimes b)=\pi_i(ba)$.)  It is easy to see
that the series $f_{\alpha,h}=\sum\limits_{m\in\Z}
x^\alpha(q^mz_i/z_j) e^{mh(\alpha)}$ converges; it can be expressed
via theta-functions, though we won't need it.

Similarly, one shows that the second term in (3.1) equals

$$ -\sum\limits_{\alpha\in\Delta^+}
{\sum\limits_{m\in\Z,j}}'x^{-\alpha}(q^mz_i/z_j)e^{-mh(\alpha)}
\pi_i\otimes \pi_j(e_\alpha\otimes
f_\alpha)\Cal F.\tag 3.4$$

As for the third summand, one must be more careful, since $J^\pm_h(z)$
contains zero modes, and therefore we have:
$\operatornamewithlimits{lim}\limits_{m\to +\infty} q^{\pm m}z
J^\pm_{h} (q^{\pm m}z) =\pm\frac{1}{2}h\otimes t^0$. This gives us the
following expression for the third term in (3.1): $$\gather
-\sum\limits_{l=1}^r
\text{V.P.}{\sum\limits_{m\in\Z, j}}' x^0(q^mz_i/z_j)
\pi_i\otimes \pi_j(x_l\otimes x_l)\Cal F\\
+\sum\limits_{l=1}^r \Tr (\Phi^1\ldots
\Phi^nq^{-\d}e^h x_l),\tag 3.5\endgather$$
where we define the principal value of a series as
$\text{V.P.}\sum\limits_{m\in\Z}a_m=
\operatornamewithlimits{lim}\limits_{M\to+\infty}\sum\limits_{m=-M}^{M}a_m$.

Finally, we obtain the following theorem.

\proclaim{Theorem 3.1} The correlation function (2.6) satisfies the
following system of differential equations:

$$\split (k+1)z_i\frac{\d}{\d z_i} \Cal F =
&-\sum\limits_{\alpha\in\Delta^+}\sum\limits_{j\ne i} f_{\alpha, h}(z_i/z_j)
\pi_i\otimes\pi_j (f_\alpha\otimes e_\alpha) \Phi \\
&-\sum\limits_{\alpha\in\Delta^+}\sum\limits_{j\ne i} f_{-\alpha, h}(z_i/z_j)
\pi_i\otimes\pi_j (e_\alpha\otimes f_\alpha) \Phi \\
&-\sum\limits_{l=1}^r\sum\limits_{j\ne i} f_{0}(z_i/z_j)
\pi_i\otimes \pi_j
(x_l\otimes x_l) \Phi \\ &-\pi_i(\Omega_h)\Cal F
+\sum\limits_{l=1}^r \pi_i(x_l)\frac{\d}{\d x_l}\Cal F,\endsplit\tag
3.6$$
where the functions $f_{\alpha, h}$ are defined as follows:
$$f_{\alpha, h}(t)=\sum\limits_{m\in\Z} x^\alpha (q^mt)e^{mh(\alpha)},
\quad \alpha\in\Delta^+$$ $$f_{-\alpha, h}(t)=\sum\limits_{m\in\Z}
x^{-\alpha} (q^mt)e^{-mh(\alpha)}, \quad \alpha\in\Delta^+$$
$$f_0(t)=\text{V.P.}\sum\limits_{m\in\Z} x^0(q^mt),\tag 3.7$$
and the operator $\Omega_h$ equals
$$\split
\Omega_h=\text{V.P.}\sum\limits_{m\ne 0}
&\left( \sum\limits_{\alpha\in\Delta^+}
(q^{-|\alpha|}e^{h(\alpha)})^m\frac{q^{mg}}{1-q^{mg}}e_\alpha f_\alpha\right.\\
&+\sum\limits_{\alpha\in\Delta^+}
(q^{|\alpha|}e^{-h(\alpha)})^m\frac{1}{1-q^{mg}}f_\alpha e_\alpha\\
&+\left.\sum\limits_{l=1}^r
\frac{1+q^{mg}}{2(1-q^{mg})}x_l^2\right)\endsplit\tag 3.8$$
\endproclaim

The functions $f_{\alpha,h}$ can be easily written in terms of
the theta-functions with characteristics. We do not give these
expressions here since we are not going to use it in this paper.

Let us look at  equations (3.6) more closely. This system
looks very much like  some (twisted) elliptic version of
Knizhnik-Zamolodchikov (KZ)
system.  The only term which breaks this analogy is $\sum
\pi_i(x_l)\frac{\d}{\d x_l}\Cal F$. Due to this term, this system of
equations is not closed, since we have no equations which would allow
us to find the derivatives $\frac{\d}{\d x_l}\Cal F$.

Let us move further. Note that our correlation function $\Cal F$ depends
not only on $z_i$ but also on the modular parameter $q$. Let us derive
the equations for $\d _q\Cal F$.

Obviously,

$$q\frac{\d}{\d q}\Cal F= \Tr (\Phi^1\ldots \Phi^n q^{-\d}e^h(-\d))\tag
3.9$$

Now let us substitute into this equation the expression for $\d$ given
by the Sugawara construction (1.3). This yields:

$$(k+1)q\frac{\d}{\d q}\Cal F =
\sum\limits_{m\in\Z}\biggl( \sum\limits_{\alpha\in\Delta^+}
a_{m,\alpha}+\frac{1}{2} \sum\limits_{l=1}^r b_{m,l}\biggr), \tag
3.10$$
where
$$a_{m,\alpha}=\Tr (\Phi^1\ldots \Phi^n
q^{-\d}e^h:e_\alpha[|\alpha| +mg]f_\alpha
[-|\alpha|-mg]:)$$ $$b_{m,l}=\Tr (\Phi^1\ldots \Phi^nq^{-\d}e^h
:x_l[mg]x_l[-mg]:)$$

Let us calculate $a_{m,\alpha}$ and $b_{m,l}$. Consider first case $m
\ge 0$. Then, using the defining commutation relation for intertwiners
(2.1), we get

$$\split a_{m,\alpha}=&\Tr (\Phi^1\ldots \Phi^nq^{-\d}e^h
f_\alpha[-|\alpha|-mg]e_\alpha[|\alpha|+mg]) \\
=&\sum\limits_{j}z_j^{|\alpha|+mg}\Tr (\Phi^1\ldots \Phi^j_{e_\alpha
u_j}(z_j)\ldots \Phi^nq^{-\d}e^h f_\alpha[-|\alpha|-mg])\\ &\qquad +
\Tr (\Phi^1\ldots
\Phi^ne_\alpha[|\alpha|+mg]q^{-\d}e^hf_\alpha
[-|\alpha|-mg]).\endsplit$$

Since $e_\alpha[l] q^{-\d}e^h=q^l e^{-h(\alpha)}q^{-\d}e^h
e_\alpha[l]$, and $[e_\alpha [l], f_\alpha
[-l]]=h_\alpha + lk$, we get

$$\split a_{m,\alpha}= &q^{|\alpha|+mg}e^{-h(\alpha)}a_{m,\alpha}\\
&+\sum\limits_jz_j^{|\alpha|+mg}\Tr (\Phi^1\ldots\Phi^j_{e_\alpha
u_j}(z_j) \ldots \Phi^nq^{-\d}e^hf_\alpha[-|\alpha|-mg])\\
&+q^{|\alpha|+mg}e^{-h(\alpha)}\left(k(|\alpha|+mg)+\frac{\d}{\d
h_\alpha}\right)\Cal F\endsplit$$

Therefore, denoting $q^{|\alpha|+mg}e^{-h(\alpha)}=c$, we see that for
$m\ge 0$ $$\gather a_{m,\alpha} =\frac{1}{1-c}\left( \sum\limits_j
z_j^{|\alpha|+mg} \Tr (\Phi^1\ldots \Phi^j_{e_\alpha u_j}(z_j)
\ldots \Phi^nq^{-\d}e^h f_\alpha[-|\alpha|-mg])\right.\\ +\left.c
\left(k(|\alpha|+mg) +\frac{\d}{\d h_\alpha}\right)\Cal F.\right).
\endgather$$

Similar arguments show that
$$\Tr (\Phi^1\ldots \Phi^j_{e_\alpha
u_j}(z_j) \ldots \Phi^nq^{-\d}e^h f_\alpha[-|\alpha|-mg])=
\frac{1}{1-c^{-1}} \sum\limits_i z_i^{-|\alpha|-mg}\pi_i\otimes \pi_j
(f_\alpha\otimes e_\alpha)\Cal F.$$

This gives us the final answer: for $m \ge 0$,

$$a_{m,\alpha}=\left(
-\frac{c}{(1-c)^2}\sum\limits_{i,j}
(z_j/z_i)^{|\alpha|+mg}\pi_i\otimes \pi_j(f_\alpha\otimes e_\alpha)
+\frac{c}{1-c}\left(k(|\alpha|+mg)+\frac{\d}{\d
h_\alpha}\right)\right)\Cal F,\tag 3.11$$
where $c=q^{|\alpha|+mg}e^{-h(\alpha)}$.

Similar considerations for $m<0$ show that $$a_{m,\alpha} =\left(
-\frac{c}{(1-c)^2}\sum\limits_{i,j}
(z_j/z_i)^{|\alpha|+mg}	\pi_i\otimes\pi_j (f_\alpha\otimes e_\alpha)
+\frac{1}{1-c}\left(k(|\alpha|+mg)+\frac{\d}{\d
h_\alpha}\right)\right)\Cal F,\tag 3.12 $$
where $c$ is given by the expression above.

Now we change the notation: $c=q^{mg}$. Then the
expressions for $b_{m,l}$ are obtained in a similar way:

for $m>0$: $$ b_{m,l}=\left( -\frac{c}{(1-c)^2}\sum\limits_{i,j}
(z_j/z_i)^{mg}
\pi_i\otimes\pi_j(x_l\otimes x_l) + \frac{c}{1-c}kmg\right)\Cal F\tag 3.13$$

for $m<0$: $$b_{m,l}=\left( -\frac{c}{(1-c)^2} \sum\limits_{i,j}
(z_j/z_i)^{mg}
\pi_i\otimes \pi_j(x_l\otimes x_l) +\frac{1}{1-c}kmg\right)\Cal F,\tag
3.14$$

for $m=0$: $$b_{0,l}=\Tr (\Phi^1\ldots \Phi^nq^{-\d}e^hx_l^2)
=\frac{\d^2}{\d x_l^2}\Cal F.\tag 3.15$$

This completes the calculations. We can rewrite the result using the
following identity:
$$\sum\limits_{m\in\Z}\frac{(qz)^{|\alpha|+mg}e^{-h(\alpha)}}
{(1-q^{|\alpha|+mg}e^{-h(\alpha)})^2}=e^{2\pi\i\zeta|\alpha|/g}
\phi\left(\frac{h(\alpha)}{2\pi\i}-\frac{|\alpha|}{g} \tau, \zeta\right),$$
where $z=e^{2\pi\i\zeta/g}, q=e^{2\pi\i\tau/g}$ and the function $\phi$
is given by

$$\phi(x,\zeta) =\sum\limits_{m\in\Z} \frac{e^{2\pi\i m\tau}e^{-2\pi\i
x} e^{2\pi\i m \zeta }}{(1-e^{2\pi\i m \tau}e^{-2\pi\i x})^2}.$$

It is easy to check that for $0< \text{Im } \zeta < \text{Im }\tau$ we
have the following expression for $\phi(x,\zeta)$:

$$\align
\phi(x,\zeta)=& \frac{-1}{2\pi\i} \frac{\d}{\d x}
\sum\limits_{m\in\Z} \frac{e^{2\pi\i m\zeta}}{1-e^{2\pi\i m \tau}
e^{-2\pi\i x}}\\
=& -\frac{1}{4\pi}\frac{\d}{\d x} \left( \frac{\theta_1(\pi(x-\zeta))
\theta'_1(0)}{\theta_1(\pi x)\theta_1(\pi\zeta)}\right).\endalign $$
This formula defines analytic continuation of $\phi(x,\zeta)$ for all
$\zeta,x \ne m\tau+n$. Recall that by definition
$$\align
\theta_1(\zeta|\tau)=&2e^{\pi\i\tau/4}\text{sin } \zeta
\prod\limits_{n\ge 1}(1-e^{2\pi\i n\tau}e^{2 \i \zeta})
	(1-e^{2\pi\i n\tau}e^{-2 \i \zeta} )(1-e^{2\pi\i n\tau})\\
=&2\sum\limits_{n=0}^{\infty} (-1)^ne^{\pi\i\tau(n+\frac{1}{2})^2} \text{sin
}(2n+1) \zeta.\endalign$$

Similarly, one can show that
$$\sum\limits_{m\ne 0}\frac{(qz)^{mg}}{(1-q^{mg})^2}=
\operatornamewithlimits{lim}\limits_{x\to 0} (\phi(x, \zeta)+\frac{1}
{4 \text{sin}^2 \pi x}).$$
Expanding the right-hand side of this equation in Laurent series
around $x=0$, we get that
$$\sum\limits_{m\ne 0}\frac{(qz)^{mg}}{(1-q^{mg})^2}=
\frac{\theta''_1(\pi\zeta)}{8\theta_1(\pi\zeta)}-
\frac{\theta'''_1(0)}{24\theta'_1(0)}+
\frac{1}{12}.$$

Finally, we can formulate the result in the following form:

\proclaim{Theorem 3.2} The correlation function $\Cal F$ satisfies the
following differential equation:

$$\split (k+1)q\frac{\d}{\d q} \Cal F =& \sum\limits_{\alpha\in\Delta^+}
\biggl(
-\sum\limits_{i,j} e^{2\pi\i |\alpha|(\zeta_j-\zeta_i)/g}
    \phi\left(\frac{h(\alpha)}{2\pi\i}-\frac{|\alpha|}{g} \tau,
	\zeta_j-\zeta_i\right)
\pi_i\otimes\pi_j(f_\alpha\otimes e_\alpha)\biggr.\\
 &\qquad + \biggl. u_{\alpha,h}\frac{\d}{\d h_\alpha}+
v_{\alpha,h}k\biggr)\Cal F\\
&+\frac{1}{2}\sum\limits_{l=1}^r\left(-\sum\limits_{i,j}
\phi_0(\zeta_j-\zeta_i)\pi_i\otimes\pi_j(x_l\otimes x_l) +
v_{0}k\right)\Cal F\\
&+\frac{1}{2}\Delta_{\frak h}\Cal F,
\endsplit\tag 3.16   $$
where $z_i=e^{2\pi\i\zeta_i/g}, q=e^{2\pi\i\tau/g}$ and
$$\align
\phi(x,\zeta)=&\sum\limits_{m\in\Z} \frac{e^{2\pi\i m\tau}e^{-2\pi\i
x} e^{2\pi\i m \zeta }}{(1-e^{2\pi\i m \tau}e^{-2\pi\i x})^2}\\
=&-\frac{1}{4\pi}\frac{\d}{\d x} \left( \frac{\theta_1(\pi(x-\zeta))
\theta'_1(0)}{\theta_1(\pi x)\theta_1(\pi\zeta)}\right)\endalign$$

$$\phi_0(\zeta)=\sum\limits_{m\ne 0}\frac{(qz)^{mg}}{(1-q^{mg})^2}
=\frac{\theta''_1(\pi\zeta)}{8\theta_1(\pi\zeta)}-
\frac{\theta'''_1(0)}{24\theta'_1(0)}+
\frac{1}{12}$$

$$u_{\alpha,h}=\sum\limits_{m\ge
0}\frac{q^{|\alpha|+mg}e^{-h(\alpha)}}{1-q^{|\alpha|+mg}e^{-h(\alpha)}}
+\sum\limits_{m<0}\frac{1}{1-q^{|\alpha|+mg}e^{-h(\alpha)}}\tag 3.17$$

$$v_{\alpha,h}=\sum\limits_{m\ge
0}\frac{q^{|\alpha|+mg}e^{-h(\alpha)}}{1-q^{|\alpha|+mg}e^{-h(\alpha)}}
(|\alpha|+mg)
+\sum\limits_{m<0}\frac{1}{1-q^{|\alpha|+mg}e^{-h(\alpha)}}(|\alpha|+mg)$$

$$v_0=\sum\limits_{m> 0}\frac{q^{mg}}{1-q^{mg}}mg
+\sum\limits_{m<0}\frac{1}{1-q^{mg}}mg$$

\endproclaim

\demo{Remark 1} We could as well write $u_{\alpha,h}, v_{\alpha,h}$ in
terms of theta-functions, but it is not necessary: in the next section
we show that the terms containing $u_{\alpha,h}, v_{\alpha,h}$ can be
cancelled by a suitable renormalization of $\Cal F$.\enddemo

\demo{Remark 2} Note that the function $\frac{\theta_1(\pi(x-\zeta))
\theta'_1(0)}{\theta_1(\pi \zeta)\theta_1(\pi x)}$ plays a very
special role in this whole story. The same function also appears later
in the integral formulas for $\Cal F$ in the case $\g=\sltwo$ (see
section 5 below); it also appears in the expressions for the functions
$f_{\alpha,h}(z)$ (cf. Theorem 3.1) for $\g=\sltwo$. \enddemo

Let us summarize the results obtained so far. We consider $\Cal F$ as a
function of $z_1\ldots z_n,q$ with values in $V=V_1\otimes\ldots
\otimes V_n\otimes C^{\infty}(\frak h_\R)$. Then theorems 3.1 and 3.2 give us
differential equations for $\Cal F$ of the following form:
$$z_i\frac{\d}{\d z_i}\Cal F = A_i(z_1\ldots z_n, q)\Cal F$$
$$q\frac{\d}{\d q}\Cal F = B(z_1\ldots z_n,q) \Cal F,\tag 3.18$$ where
$A_i, B$ are some operators in $V$ which are defined for $|z_1|>\ldots
|z_n|>|qz_1|$.

Note also that we can easily find asymptotics of $\Cal F$ as $q\to 0$.
Indeed, it is easy to show that as $q\to 0$,

$$\split
\Cal F\sim& <v_{\lambda,k}, \Phi^1\ldots\Phi^n q^{-\d}e^h
v_{\lambda,k}>\\
=&e^{h(\lambda)}q^{\frac{<\lambda,\lambda>}{2(k+1)}}\Psi\endsplit\tag
3.19$$
where $\Psi=<v_{\lambda,k}, \Phi^1(z_1)\ldots \Phi^n(z_n)
v_{\lambda,k}>$ is the ``correlation function on the sphere'' with
values in $V_1\otimes\ldots \otimes V_n$. For the case when $V_i$ are
highest or lowest weight representations, which is our main example, this
function is very well studied; in particular, it is known that it is
well defined for $|z_1|>\ldots|z_n|$ and satisfies trigonometric KZ
equation in this region (see \cite{FR} and references therein); there
are integral formulas for this function (see \cite{SV},\cite{Ch}). So
we consider $\Psi$ as a well-known function.

Now standard arguments from the theory of ordinary differential
equations show that the equation (3.18) together with boundary
condition (3.19) uniquely determine $\Cal F$; in fact, we only need the
equation for $q\frac{\d}{\d q}\Cal F$.

\demo{Remark} Note that we never used the fact that $M_{\lambda_i,k}$
are Verma modules. In fact, all the statements of this section are
still valid if we consider any modules from the category $\Cal O$
instead of Verma modules -- provided that the intertwiners exist.
\enddemo

\heading{\bf 4. Connection with the heat equation}\endheading

In this section we will show, following the ideas of Bernard
\cite{Ber} that the equation (3.16) is closely related to the heat
equation on $\frak h_\R$. This is of special interest for
$n=0$ and $n=1$. In this cases  $\Cal F$ does not depend on $z_i$, so (3.16) is
the only non-trivial equation  for $\Cal F$.

Let us denote $\Cal F_0=\Tr |_{M_{0,k}}(q^{-\d} e^h)$. It can be easily
calculated explicitly: since the character of $M_{0,k}$ coincides with
the character of $U\gtwisted^-$, we see that

$$\Cal F_0=\frac{1}{\prod\limits_{m>0}(1-q^{mg})^r
\prod\limits_{\alpha,m}(1-e^{-h(\alpha)}q^{|\alpha|+mg})},\tag 4.1$$
where the second product is taken over all $\alpha\in\Delta, m\in\Z$
such that $|\alpha|+mg>0$. This can be rewritten as follows:

$$\frac{1}{\Cal F_0}=\prod\limits_{m>0}(1-q^{mg})^r
\prod\limits_{\alpha\in\Delta^+}\left(\prod\limits_{m\ge
0}(1-e^{-h(\alpha)}q^{|\alpha|+mg})\prod\limits_{m<0}(1-e^{h(\alpha)}
q^{-|\alpha|-mg}) \right).\tag 4.2$$

Explicit calculation shows that $$\split q\frac{\d}{\d q}
\frac{1}{\Cal F_0}=& -\sum\limits_{\alpha\in \Delta^+}
\left(\sum\limits_{m\ge
0}\frac{q^{|\alpha|+mg}e^{-h(\alpha)}}{1-q^{|\alpha|+mg}e^{-h(\alpha)}}
(|\alpha|+mg)
+\sum\limits_{m<0}\frac{1}{1-q^{|\alpha|+mg}e^{-h(\alpha)}}(|\alpha|+mg)
\right)\frac{1}{\Cal F_0}\\
&\qquad -r\sum\limits_{m>0}
\frac{q^{mg}}{1-q^{mg}}mg\frac{1}{\Cal F_0}\\
=&\left(-\sum\limits_{\alpha\in\Delta^+}v_{\alpha,h} - \frac{r}{2}
v_0\right) \frac{1}{\Cal F_0},\endsplit\tag 4.3$$
where $v_{\alpha,h}, v_0$ are defined in Theorem 3.2.

Similarly, if $x\in\frak h$, $$ \frac{\d}{\d
x}\frac{1}{\Cal F_0}=\left(\sum\limits_{\alpha\in\Delta^+}
x(\alpha)u_{\alpha, h}\right)\frac{1}{\Cal F_0}\tag 4.4$$
where $u_{\alpha,h}$ is defined in Theorem 3.2.

Now let us come back to Theorem 3.2, which we rewrite in the
following form $$\split q\frac{\d}{\d q}\Cal F =& \frac{1}{k+1}A\Cal F +
\sum\limits_{\alpha\in\Delta^+}
\left(\frac{1}{k+1}u_{\alpha,h}\frac{\d}{\d
h_\alpha}+\frac{k}{k+1}v_{\alpha,h}\right) \Cal F\\
&+\frac{k}{k+1}\frac{r}{2}v_0\Cal F\\ &+\frac{1}{2(k+1)}\Delta_{\frak
h}\Cal F,\endsplit\tag 4.5$$
where we for brevity denoted
$$\split
A=&-\sum\limits_{\alpha\in\Delta^+}
\sum\limits_{i,j}e^{2\pi\i|\alpha|/g}
 \phi\left(\frac{h(\alpha)}{2\pi\i}-\frac{|\alpha|}{g} \tau,
	\zeta_j-\zeta_i\right)
\pi_i\otimes\pi_j(f_\alpha\otimes
e_\alpha)\\ &-\frac{1}{2}\sum\limits_{l=1}^r\sum\limits_{i,j}
\phi_0(\zeta_j-\zeta_i)\pi_i\otimes\pi_j(x_l\otimes x_l),\endsplit\tag 4.6$$
where $\phi(x,\zeta),\phi_0(\zeta)$ are defined in Theorem 3.2.
Comparing (4.5) with (4.4) and (4.3), we get the following
statement.
\proclaim{Proposition 4.1}
$$q\frac{\d}{\d q}\left(\frac{\Cal F}{\Cal F_0}\right)=
\frac{1}{k+1}\left((A\Cal F)\frac{1}{\Cal F_0}+(q\frac{\d}{\d
q} \frac{1}{\Cal F_0}) \Cal F + \sum\limits_{l=1}^r\frac{\d \Cal F}{\d x_l}
\frac{\d (1/\Cal F_0)}{\d x_l} + \frac{1}{2}(\Delta_{\frak
h}\Cal F)\frac{1}{\Cal F_0}\right) \tag 4.7$$
\endproclaim

\demo{Proof} The only not immediately obvious step in the proof is
dealing with the terms containing $u_{\alpha,h}$. As for them, $$
\left(\sum\limits_{\alpha\in\Delta^+} u_{\alpha,h}\frac{\d \Cal F}{\d
H_\alpha}\right)\frac{1}{\Cal F_0}=\left(\sum\limits_{\alpha\in\Delta^+}
u_{\alpha,h} <h_{\alpha}, \text{grad }\Cal F>
\right)\frac{1}{\Cal F_0}.$$

Comparing it with (4.4), we see that this equals to

$$<\text{grad }\Cal F,\text{grad }\frac{1}{\Cal F_0}>= \sum\limits_l
\frac{\d \Cal F}{\d x_l} \frac{\d (1/\Cal F_0)}{\d x_l}.$$
\qed
\enddemo

\proclaim{Corollary}
$$\left(q\frac{\d}{\d q}-\frac{1}{2}\Delta_{\frak h}\right)
\frac{1}{\Cal F_0}=0. \tag 4.8$$
\endproclaim
\demo{Proof} Rewrite (4.7) in the form
$$\left( q\frac{\d}{\d q} -\frac{1}{2(k+1)}\Delta_{\frak h}\right)
\left(\frac{\Cal F}{\Cal F_0}\right)=
\frac{1}{k+1}(A\Cal F)\frac{1}{\Cal F_0}+\frac{1}{k+1}\Cal F\left(
q\frac{\d}{\d q}-\frac{1}{2}\Delta_{\frak h}\right)\frac{1}{\Cal F_0}$$
and substitute there $\Cal F=\Cal F_0 = \Tr|_{M_{0,k}}(q^{-\d}e^h)$.
\enddemo

So, finally we have the following theorem:
\proclaim{Theorem 4.1}

$$\align
&\left( q\frac{\d}{\d q} -\frac{1}{2(k+1)}\Delta_{\frak
h}\right)\frac{\Cal F}{\Cal F_0} =\\
-\frac{1}{k+1}
\sum\limits_{i,j}&\left(  \sum\limits_{\alpha\in\Delta^+}
    e^{2\pi\i|\alpha|/g}
	\phi\left(\frac{h(\alpha)}{2\pi\i}-\frac{|\alpha|}{g} \tau,
	\zeta_j-\zeta_i\right)
    \pi_i\otimes\pi_j(f_\alpha\otimes e_\alpha)\right.\\
  +&\quad\left.\frac{1}{2}\sum\limits_{l=1}^{r}
\phi_0(\zeta_j-\zeta_i)\pi_i\otimes\pi_j(x_l\otimes x_l)\right)
\frac{\Cal F}{\Cal F_0},\tag
4.9\endalign$$
where $\phi(x,\zeta), \phi_0(\zeta)$ are defined in Theorem 3.2.
\endproclaim

\proclaim{Corollary}Let  $n=0$, i.e. $\Cal F=\Tr (q^{-\d}e^h)$. Then
$\Cal F/\Cal F_0$ satisfies the heat equation:
$$\left( q\frac{\d}{\d q} -\frac{1}{2(k+1)}\Delta_{\frak
h}\right)\frac{\Cal F}{\Cal F_0} = 0.\tag 4.10$$
\endproclaim

This equation (for
untwisted algebra) was first derived by Bernard (\cite{Ber}).

Let us consider more interesting example $n=1, \Cal F= \Tr|_{M_{\lambda,
k}} (\Phi^1(z)q^{-\d}e^h)$. In this case $\Cal F$ takes values in some
module $V$ over $\g$. From the weight considerations it is clear that
in fact $\Cal F$ takes values in the zero-weight subspace $V (0)$.
Therefore, the equation (4.9) reduces to
$$\left( q\frac{\d}{\d q}
-\frac{1}{2(k+1)}\Delta_{\frak h}\right)\frac{\Cal F}{\Cal F_0} =
-\frac{1}{k+1}\sum\limits_{\alpha\in\Delta^+}
   \phi\left(\frac{h(\alpha)}{2\pi\i}-\frac{|\alpha|}{g} \tau,0\right)
f_\alpha e_\alpha$$

Note that $\phi(x,0)$ is an elliptic function of a very simple form.
Indeed, from the expression
$$\phi(x,0)=\sum\limits_{m\in\Z} \frac{e^{2\pi\i m\tau}e^{-2\pi\i
x} }{(1-e^{2\pi\i m \tau}e^{-2\pi\i x})^2}$$
it is obvious that $\phi(x,0)$ is an even elliptic function with periods
$1,\tau$ and poles of order 2 at the points $m\tau +n, m,n\in\Z$.
Finding the leading coefficient at $u=0$, we see that

$$\phi(x,0)=-\frac{1}{(2\pi)^2}\left(\wp(x)+c\right),\tag 4.11$$
for some constant $c=c(q)$ (not to be confused with $c$ of section 3);
we won't need an explicit expression for it. We will, however, use two
properties of $c(q)$: as $q\to 0$, $c(q)\to 0$ and
if $q\in\R$ then $c(q)\in\R$.

This implies
\proclaim{Proposition 4.2} Let $n=1$. Then
the correlation function $\Cal F$ satisfies
the following equation:

$$\left( q\frac{\d}{\d q}-\frac{1}{2(k+1)}\Delta_{\frak h}\right)
\frac{\Cal F}{\Cal F_0}=
\frac{1}{(2\pi)^2(k+1)}\sum\limits_{\alpha\in\Delta^+}\left(\wp\left(\frac{\tau|\alpha|}{g}
-\frac{h(\alpha)}{2\pi \i}\right)+c\right)f_\alpha e_\alpha
\frac{\Cal F}{\Cal F_0} .\tag 4.12$$
\endproclaim

\heading{\bf 5. Example: $\g=\sltwo, n=1$.}\endheading

	Let us consider the simplest possible case $\g=\sltwo, n=1$. Let
$e,f,h$ be the standard basis of $\sltwo$. We can identify $\R\simeq\frak
h_{\R}$ by $x\mapsto -\pi \i x h$. Under this identification,
the Killing  form becomes $-2\pi^2x^2$, and the Laplace operator:
$\Delta_{\frak h}=-\frac{1}{2\pi^2}\frac{d^2}{d x^2}$.

Let $V_\mu, \mu\in\C$ be an irreducible
 highest-weight module over $\sltwo$ with the
highest  weight $\frac{\mu}{2}\alpha$, where $\alpha$ is the
positive root of $\sltwo$.  In other words, the action of $\frak h$ on
the highest  weight vector of $V_\mu$ is given by $hv_\mu =
\mu v_\mu$.

Note that the zero-weight subspace $V_\mu(0)$ is not empty iff
$\mu = 2m, m\in\Z_+$. For this reason, from now on we assume that
$V_\mu=V_{2m}$ is a finite-dimensional module of dimension $2m+1$.
Then $V_{2m}(0)$ is one-dimensional
and $f_\alpha e_\alpha|_{V_{2m}(0)}=m(m+1)$. Also, it is well-known
that in this case the intertwining operator $\Phi: M_{\lambda,k}\to
M_{\lambda,k}\otimes V_\mu$ exists and is unique up to a constant
factor for generic values of $\lambda, k$.

So we can rewrite (4.12)
as follows:

$$\left(-4\pi \i(k+1)\frac{\d}{\d \tau} +\frac{\d ^2}{\d
x^2}\right)F=m(m+1)\left(\wp\left(\frac{\tau}{2} +x\right)+c\right)
F,\tag 5.1$$
where $F=\Cal F/\Cal F_0$ is a $\C$-valued function of
$x\in\R,\tau\in\Cal H$, $\Cal H$ being the upper half-plane.

Note that this equation does not depend on the highest weight
$\lambda$ of the Verma module we consider: the only dependence is the
boundary conditions (3.19). Note, however, that $F$ has the following
periodicity property: $F(x+1)=e^{-\pi\i\lambda}F(x)$; in particular,
if $\lambda\in 2\Z$ then $F(x)$ is periodic:
$F(x+1)=F(x)$.

Note that in the limit $k\to -1$ the equation (5.1) becomes the Lam\'e
equation (see \cite{WW}):
$$\frac{d^2}{d x^2}F
=m(m+1)\left(\wp\left(\frac{\tau}{2}+x\right)+c\right)F.\tag 5.2$$
We discuss this relation in the next section.

To find an explicit expression for $F$, we use formula (2.5),
which allows us to rewrite $F$ in terms of untwisted intertwiners
$\hat\Phi(\zeta)$ as follows:

$$F(x,q)=\frac {\Tr |_{\Cal M_{\Lambda, K}}\left( \hat\Phi[0]
q^{-2d-\frac{h}{2}}e^{-\pi \i x h}\right)} {\Tr |_{\Cal M_{k,K}}\left(
q^{-2d -\frac{h}{2}}e^{-\pi \i x h}\right)},$$
where $K=2k, \Lambda =
\lambda +k$.

This enables us to use well-known integral formulas for
$\hat\Phi(\zeta)$. For this purpose we recall the Wakimoto realization
of $\widehat{\sltwo}$ (\cite{Wa}), following \cite{BF}.  Let us introduce the
algebra $A$, generated by $\alpha_n,\beta_n,\gamma_n, n\in\Z$ with the
relations: $$\aligned [\alpha_n,\alpha_m]&=2n\delta_{n+m,0}\\
[\beta_n,\gamma_m]&=\delta_{n+m,0}\endaligned\tag 5.3$$
and all the other commutators vanish.

Next, we can define the module $H_{\Lambda}, \Lambda\in\C$ over $A$ as
a  module
generated by a vacuum vector $v_\Lambda$ with the properties:
$$\aligned
\alpha_nv_\Lambda &=0,\quad n>0 \\
\beta_n v_\Lambda&=\gamma_{n+1}v_\Lambda=0, \quad n\ge 0\\
\alpha_0v_\Lambda& =
\frac{\Lambda}{\sqrt{\kappa}}v_\Lambda,\endaligned\tag 5.4$$
where $\kappa=K+2\ne 0$.

Define the normal ordering in $A$ by:

$$\gather :\alpha_n \alpha_m: =\cases \alpha_n\alpha_m, \quad m>0\\
\alpha_m\alpha_n, \quad m\le 0\endcases\\ :\beta_n \gamma_m: =\cases
\beta_n\gamma_m, \quad m>0\\ 			\gamma_m\beta_n, \quad
m\le 0\endcases\tag 5.5
\endgather$$

Finally, introduce the free bosonic fields  as follows:

$$\aligned
\alpha(z)=\sum\limits_{n\in\Z} \alpha_n z^{-n-1}\\
\beta(z)=\sum\limits_{n\in\Z} \beta_nz^{-n-1}\\
\gamma(z)=\sum\limits_{n\in\Z} \gamma_n z^{-n}\endaligned\tag 5.6$$

Then we have the following proposition, due to Wakimoto \cite{Wa}.
\proclaim{Proposition}{\rm cf. \cite{BF}} The following formulas give
an action of $\widehat{\sltwo}$ in $H_{\Lambda}$. Moreover, $H_\Lambda$
endowed with this action is isomorphic to Verma module $\Cal
M_{\Lambda,K}$ for generic $K, \Lambda$.

$$\aligned J_e(z)&=\beta(z)\\
J_h(z)&=-2:\beta(z)\gamma(z):+\sqrt{\kappa}\alpha(z)\\
J_f(z)&=-:\gamma^2(z)\beta(z):+\sqrt{\kappa}\alpha(z)\gamma(z)+K\gamma'(z)
\endaligned \tag 5.7$$
\endproclaim

We also need explicit formulas for the intertwiners in
Wakimoto realization. Let us introduce the operator $e^{c\alpha}\colon
H_\Lambda\to H_{\Lambda+2c\sqrt{\kappa}}$ which maps $v_\Lambda$ to
$v_{\Lambda+2c\sqrt{\kappa}}$ and commutes with all the generators of
$A$ except $\alpha_0$. Then we can introduce the vertex operators
$X(\mu), \mu\in\frak h^*$ by:

$$X(c\alpha, z)= :\text{exp}\left(c\int \alpha(z)\, dz\right):
=\text{exp}\left(c\sum\limits_{n<0}\frac{\alpha_n}{-n}z^{-n}\right)
\text{exp}\left(c\sum\limits_{n>0}\frac{\alpha_n}{-n}z^{-n}\right)
e^{c\alpha}z^{c\alpha_0}$$

and the screening operators $U(t)$:
$$U(t)=\beta(t)X(-\frac{\alpha}{\sqrt{\kappa}},t)$$

Then the intertwining operator $\hat\Phi(z):\Cal M_{\lambda, K}\to\Cal
M_{\nu,K} \otimes V_\mu$ can be written as follows (we assume that
$\mu\in\Z$, so $V_\mu$ is finite-dimensional, and
$\lambda+\mu-\nu=2m$)

$$\hat\Phi(z) v =\sum\limits_{n\ge 0}\left(\int\limits_{\Delta}
X(\frac{\mu}{2\sqrt{\kappa}}\alpha) (-\gamma(z))^n U(t_1)\ldots U(t_m)
dt_1\ldots dt_m \right) v\otimes \frac{e^n}{n!}v_{-\mu},\tag 5.8 $$
where $v_{-\mu}$ is the lowest-weight vector in $V_\mu$ and the cycle
of integration $\Delta\subset \C^m\setminus \{ t_i=0,z, t_i=t_j\}$ is
chosen so that one can define single-valued branches of $\text{log
}t_i, \text{log }(t_i-t_j), \text{log }(t_i-z)$ on $\Delta$. General
results of the homology theory  of local systems (see \cite{SV})
show that such cycle is essentially unique.

In particular, this shows that the zero-mode
component $\hat\Phi[0]: \Cal M_{\Lambda,K}\to\Cal M_{\Lambda,
K}\otimes V_{\mu}(0)$ is given by:

$$\hat\Phi[0] = \int\limits_{\Delta} X(\frac{m}{\sqrt{\kappa}}\alpha,
z) (-\gamma(z))^m U(t_1)\ldots U(t_m) dt_1\ldots dt_m\tag 5.9$$

To find $F$, we must calculate $\Tr |_{\Cal M_{\Lambda, K}}\left(
\hat\Phi[0] q^{-2d-\frac{h}{2}}e^{-\pi \i x h}\right)$. Traces of this
type can be calculated in the same way as we did in section 3; we
borrow the answer from \cite{BF, formula 3.14}, which can be rewritten
as follows:

$$\gather
\Tr |_{H_\Lambda} \left( (-\gamma(t_0))^m \prod\limits_{i=1}^m
\beta(t_i) \prod \limits_{i=0}^m X(\frac{l_i}{\sqrt{\kappa}}\alpha,
t_i) q^{-2d} e^{-\pi \i x h}\right)=\\
\text{Ch}_{H_\Lambda}\times \prod\limits_{i=0}^m
t_i^{(-l_i^2+l_i\Lambda)/\kappa}
\prod \limits_{0\le i<j\le m}\left(
\frac{\theta_1(\pi(\zeta_i-\zeta_j)|\tau)}{\theta_1'
(0|\tau)}\right)^{2l_il_j/\kappa}\times m! \prod G(t_0,
t_i|\tau,x)\tag 5.10\endgather$$

where
$$\gather q=e^{\pi \i \tau}\\ t_i=e^{2\pi \i \zeta_i}\\
\text{Ch}_{H_\Lambda} = \Tr |_{H_\Lambda}\left(q^{-2d} e^{-\pi \i x
h}\right)\\
G(t_0,t_i|\tau,x)= \sum\limits_{r\in\Z}
\frac{1}{1-q^{2r}e^{-2\pi \i x}} \frac {t_i^{r-1}}{t_0^r}=
\sum\limits_{n\in\Z} \frac{e^{2\pi \i n x}}{t_0q^{2n}-t_i}
\endgather$$

Applying this formula in our case, we get

$$\gather
\Tr |_{\Cal M_{\Lambda, K}} \left(\hat\Phi[0] q^{-2d} e^{-\pi \i  x
h}\right)=\\
\int\limits_{\Delta} \Tr|_{H_\Lambda}
\left( (-\gamma(z))^m \prod\limits_{i=1}^m
   \beta(t_i) \prod \limits_{i=1}^m
X(-\frac{\alpha}{\sqrt{\kappa}},t_i) q^{-2d} e^{-\pi
h(x+\frac{\tau}{2})}\right)\, dt_1\ldots dt_m=\\
\Tr |_{\Cal M_{\Lambda, K}} \left(q^{-2d} e^{-\pi \i  xh}\right)
z^{m(m+\Lambda)/\kappa}
\int\limits_{\Delta}
  \left( \prod\limits_{i=1}^m t_i^{(1-\Lambda)/\kappa}\right.\\
\left.\prod\limits_{i=1}^m
\left(\frac{\theta_1(\pi(\zeta_0-\zeta_i)|\tau)}
{\theta_1'(0|\tau)}\right)^{-2m/\kappa} 	 \prod\limits_{1\le
i<j\le m}
\left(\frac{\theta_1(\pi(\zeta_i-\zeta_j)|\tau)}
{\theta_1'(0|\tau)}\right)^{2/\kappa} 	m! \prod \limits_{i=1}^m
G(z,t_i|\tau, x+\frac{\tau}{2}) \right) dt_1\ldots dt_m\endgather$$

Since

$$\align
\Tr |_{\Cal M_{\Lambda, K}} \left(q^{-2d-\frac{h}{2}} e^{-\pi \i
	hx}\right)=& e^{-\pi \i (\Lambda
-k)x}q^{\frac{\lambda^2}{4(k+1)}}\Tr |_{\Cal M_{k, K}}
\left(q^{-2d-\frac{h}{2}} e^{-\pi \i hx}\right)\\ =&e^{-\pi \i
x\lambda +\pi \i \tau \frac{\lambda^2}{2\kappa}}
\Tr |_{\Cal M_{k, K}}
   \left(q^{-2d-\frac{h}{2}} e^{-\pi \i hx}\right)	\endalign$$ we
finally get

$$\gather F(x,q)= e^{-\pi \i x\lambda +\pi \i \tau
\frac{\lambda^2}{2\kappa}} z^{m(\Lambda-m)/\kappa}\\
\int\limits_{\Delta}
  \left( \prod\limits_{i=1}^m t_i^{(-1-\Lambda)/\kappa}
\prod\limits_{i=1}^m
\left(\frac{\theta_1(\pi(\zeta_0-\zeta_i)|\tau)}
{\theta_1'(0|\tau)}\right)^{-2m/\kappa}\right.\\
\left.\prod\limits_{1\le i<j\le m}
\left(\frac{\theta_1(\pi(\zeta_i-\zeta_j)|\tau)}
{\theta_1'(0|\tau)}\right)^{2/\kappa} 	m! \prod \limits_{i=1}^m
G(z,t_i|\tau, x+\frac{\tau}{2}) \right)dt_1\ldots dt_m\tag
5.11\endgather$$

Since we know that this function in fact does not depend on  $z$
(though it is not immediately obvious from the formula above), we can
put $z=1$. In this case we can use the following formula:
$$G(1, e^{2\pi \i\zeta}|\tau, x)=\frac{\i}{2} e^{-2\pi\i\zeta}
\frac{\theta_1(\pi(x-\zeta))\theta'_1(0)}{\theta_1(\pi x)\theta_1(\pi
\zeta)}, \tag 5.12$$

So we have proved the following theorem:
\proclaim{Theorem 5.1} For any $\lambda,\kappa\in\C, \kappa\ne 0,m\in\Z_+$, the
function $F_{\lambda,\kappa}(x,q)$ given by

$$\gather F_{\lambda,\kappa}(x,q)= e^{-\pi \i x\lambda +\pi \i \tau
\frac{\lambda^2}{2\kappa}}\\
\int\limits_{\Delta}
  \left( \prod\limits_{i=1}^m
     e^{2\pi\i\zeta_i(-1-\frac{\lambda}{\kappa})}
     l(\zeta_i|\tau)^{-2m/\kappa}\right.\\
\left.\prod\limits_{1\le i<j\le m}
     l(\zeta_i-\zeta_j|\tau)^{2/\kappa}
     \prod \limits_{i=1}^m
	\frac{\theta_1(\pi(x-\zeta_i+\frac{\tau}{2}))\theta'_1(0)}
		{\theta_1(\pi (x+\frac{\tau}{2})\theta_1(\pi\zeta_i)}
\right)d\zeta_1\ldots d\zeta_m,\tag 5.13\endgather$$
where
$$\gather
l(\zeta|\tau)=\frac{\theta_1(\pi \zeta|\tau)}{\theta'_1(0|\tau)}\\
q=e^{\pi \i\tau}\endgather$$
satisfies the partial differential equation

$$\left(-2\pi \i \kappa\frac{\d}{\d \tau} +\frac{\d ^2}{\d
x^2}\right)F_{\lambda,\kappa}=
m(m+1)\left(\wp\left(\frac{\tau}{2} +x\right)+c\right)
F_{\lambda,\kappa},\tag 5.14$$
and the following boundary conditions when $q\to 0$,

$$F_{\lambda, \kappa}(x,q)\sim  C(\lambda,\kappa) e^{-\pi \i x\lambda +\pi \i
\tau
\lambda^2/2\kappa},\tag 5.15$$
where

$$\gather C(\lambda,\kappa)=\int\limits_{\Delta}
  \left( \prod\limits_{i=1}^m
     e^{-2\pi\i\zeta_i \lambda/\kappa}
     (\text{sin }\pi \zeta_i)^{-2m/\kappa}\right.\\
\left.\prod\limits_{1\le i<j\le m}
     (\text{sin } \pi (\zeta_i-\zeta_j))^{2/\kappa}
     \prod \limits_{i=1}^m
	\frac{1}{ (1-e^{2\pi\i\zeta_i})}
\right)d\zeta_1\ldots d\zeta_m.\tag 5.16\endgather$$

\endproclaim

\demo{Remark} The function $F(x,q)$ in the theorem above differs from
the function $F(x,q)$ given by (5.11) by a constant factor.\enddemo

Note that in the limit $q\to 0$ the right-hand side of equation (5.14)
tends to zero and the equation itself tends to the heat equation:

$$(-2\pi\i\kappa \frac{\d}{\d\tau}+\frac{\d^2}{\d x^2})f(x,\tau)=0,\tag
5.17$$
and the asymptotics (5.15) of the functions $F_{\lambda,\kappa}$ given
by Theorem 5.1
form a basis in the space of solutions of the heat equation.
This allows us to find all the solutions of
equation (5.14) having good asymptotics as $\tau\to \i\infty$. Let us
consider the case then $\tau=\i t, t\in\R_+$ and $\kappa\in\i \R$. In
this case equation (5.14) can be considered as non-stationary
Schr\"odinger equation:

$$\frac{\d f}{\d t}= \frac{1}{2\pi\kappa} Hf,\tag 5.18$$
with the Hamiltonian
$$H= \frac{d^2}{dx^2}-m(m+1)\left(\wp(x+\frac{\tau}{2})+c(\tau)\right).
\tag 5.19$$
Note that this Hamiltonian is self-adjoint with respect to the $L^2$
norm.

In the limit $q\to 0$, (5.18) tends to
$$\left(\frac{\d }{\d t} - \frac{1}{2\pi\kappa}\frac{\d^2}{\d x^2}
\right) f(x,\tau)=0\tag 5.20$$

\proclaim{Theorem 5.2} Let us fix $\kappa\in\i\R$ and asssume that
$\tau=\i t,t\in\R$.  Let $f(x,\tau)$ be a solution of (5.18) such that

$$f(x,\tau)= f_0(x,\tau)(1+g(x,\tau))$$
where $f_0(x,\tau)$ is a
solution of (5.20) and $\operatornamewithlimits{sup}\limits_{x\in\R}
|g(x,\tau)|\to 0$ as $q\to 0 (t\to +\infty)$.
Also, let us assume that for some $\tau_0$
$f_0(x,\tau_0)$ is a rapidly decreasing function of $x$.

Then  $f(x,\tau)$ can be presented in the following
form:

$$f(x,\tau)=\int\limits_{\R}
\frac{F_{\lambda,\kappa}(x,\tau)}{C(\lambda,\kappa)}\rho(\lambda)\,
d\lambda\tag 5.21$$
with the function $\rho(\lambda)$ given by
$$\rho(\lambda)=\frac{1}{2} \int\limits_{\R} e^{\pi\i\lambda x} f_0(x,0) \,
d x$$

\endproclaim

\demo{Proof} The fact that Hamiltonian (5.19) is self-adjoint with
respect to the $L^2$ norm implies that if $f(x,\tau)$ is a solution of
Schr\"odinger equation (5.18) then $|f(x,\tau)|_{L^2(\R)}$ does not
depend on $\tau$. The same applies to $f_0$.
Next, $f_0$ -- as any solution of (5.20) -- can be written in the form

$$f_0(x,\tau)=\int\limits_{\R} e^{-\pi\i\lambda x
+\pi\i\tau\lambda^2/2\kappa}\rho
(\lambda)\, d\lambda$$
for some density function $\rho$. Letting $\tau\to 0$ and applying
inverse Fourier transform, we see that

$$\rho(\lambda)=\frac{1}{2} \int\limits_{\R} e^{\pi\i\lambda x} f_0(x,0) \,
d x$$
Now let us define the function $\tilde f$ by

$$\tilde f(x,\tau)= \int\limits_{\R}
\frac{F_{\lambda,\kappa}(x,\tau)}{C(\lambda,\kappa)}
\rho(\lambda)\, d\lambda.$$

Then Theorem (5.1) implies that $\tilde f(x,\tau)$ satisfies
Schr\"odinger equation (5.18). Moreover, it is easy to check that
convergence of $F_{\lambda,k}$ to asymptotics (5.15) is in fact
uniform in $x$. Therefore, as $q\to 0$ $\tilde f(x,\tau)$ can be
written in the form
$$\tilde f(x,\tau)=f_0(x,\tau)(1+\tilde g(x,\tau))$$
with the same function $f_0$ as above and $\tilde g$ satisfying
$\operatornamewithlimits{sup}\limits_{x\in\R} |g(x,\tau)| \to 0$
as $q\to 0$. Comparing this with the assumption of the theorem and
taking into account that the $L^2$ norms $|f(x,\tau)-\tilde
f(x,\tau)|, |f_0(x,\tau)|$ do not depend on $\tau$, we see that
$f(x,\tau)=\tilde f(x,\tau)$, which completes the proof.\qed
\enddemo
\heading {\bf 6. Lam\'e functions.}\endheading

In this section we use the theorem proved in the previous section
to find explicit formulas for Lam\'e functions. The basic idea is to let
$k\to -1$; obviously, in this limit the equation (5.1) becomes Lam\'e
equation. Unfortunately, we can not just repeat all the arguments of
the previous sections for $k=-1$, since in this case (critical level)
the whole theory fails; we can not even claim  the existence of the
intertwiners between Verma modules and evaluation representations.
Instead, we can use the integral formula (5.13) for $F$ and then find
the asymptotics  when $k\to -1$.

We illustrate this general idea on the simplest example $m=1$, when
all the calculations can be done absolutely explicitly.  In this case
the expression for $F$ given at the end of previous section reduces
to:

$$F(x,q)=e^{-\pi \i \lambda x +\pi \i\tau \lambda^2/2\kappa}
\int\limits_{\Delta}e^{2\pi \i \zeta(-1-\frac{\lambda}{\kappa})}
\left( \frac{\theta_1(\pi\zeta|\tau)}{\theta'_1(0|\tau)}\right)^{-2/\kappa}
	\frac{\theta_1(\pi(x-\zeta+\frac{\tau}{2})\theta'_1(0)}
	{\theta_1(\pi(x+\frac{\tau}{2}))\theta_1(\pi\zeta)}
\, d\zeta\tag 6.1$$

	In this case one can take $\Delta$ to be the Pochhammer loop
in the $t$-plane, i.e. the element $a_1\circ a_0 \circ a_1^{-1}
\circ a_0^{-1}$ of $\pi_1(\C\setminus \{0,1\}$, where
$a^{\pm 1}_{0,1}:[0,1]\to\C$ is the loop going around $0,1$ in the
anticlockwise/clockwise direction. This cycle is the same as the cycle
$a_0 \circ a_1^{-1}$ in the $\zeta$-plane.

Now, let us assume that $k$ approaches $-1$ along the real line from
below: $k\in\R, k<-1$. Then $\kappa<0$ and therefore $-2/\kappa >0$.
In this case we can use the saddle-point method for finding
asymptotics of the integral (6.1) (see, for example, E.T. Copson,
``Asymptotic expansions'', Cambridge Univ. Press, 1965).
\proclaim{Proposition} Let $f(\nu)$ be defined by the integral
$$f(\nu)=\int\limits_{\Delta} e^{\nu w(z)}\phi(z)\, dz, \tag 6.2$$
where $\Delta$ is a contour in $\C$, $w,\phi$ are analytic functions
on $\Delta$. Let us assume that there exists $z_0\in\Delta$ such
that:

1. $w'(z_0)=0, w''(z_0)\ne 0$

2. $\phi(z_0)\ne 0$

3. For any $z\in\Delta, z\ne z_0$ we have $|e^{w(z)}|<|e^{w(z_0)}|$.

Then $f(\nu)$ has the following asymptotics as $\nu\to +\infty$:

$$f(\nu)\sim \phi(z_0)e^{\nu w(z_0)}
\left(\frac{-2\pi}{\nu w''(z_0)}\right)^{\frac{1}{2}}.\tag 6.3$$

Moreover, if $\phi, w$ also depend analytically on some parameters
$x_i$ then the asymptotics (6.3) can be differentiated to get
asymptotics for $\frac{\d}{\d x_i} f(\nu)$.
\endproclaim

Therefore, to find the asymptotics of the integral (6.1) we must find
critical points of $ e^{\pi \i \lambda\zeta} \theta_1(\pi\zeta)$,
i.e.  points $\zeta_0$ satisfying $$ \lambda=\i
\frac{\theta'_1(\pi\zeta_0)}{\theta_1(\pi\zeta_0)}\tag 6.4$$

Of course, it is hopeless to find explicit expression for $\zeta_0$ as
a function of $\lambda$; however, for some special values of $\lambda$
-- for example, $\lambda\in \Z$ -- it can be done; we return to
this later.  Therefore, it is better to consider $\zeta_0$ as given
and define $\lambda$ by (6.4).

Let us assume the following:

1. $\theta''_1(\pi \zeta_0) \ne 0$

2. The contour $\Delta$, described above, can be deformed so that it
passes through $\zeta_0$ precisely twice, and $|e^{\pi \i
\lambda\zeta}
\theta_1(\pi\zeta)|$ attains its absolute maximum on $\Delta$ at
$\zeta_0$.

In this situation we can apply the saddle-point method, which gives us
the following asymptotics for $F$ as $k\to -1$:
$$F(x,q)\sim(1-e^{4\pi\i/\kappa})
(A(q))^{-2/\kappa}\sqrt{-\kappa/2}\times B(x,q),\tag 6.5$$

where
$$\gather A(q)=e^{-\pi \i (\lambda+\tau\lambda^2/4)}
\frac{\theta_1(\pi\zeta_0)}{\theta'_1(0)}\\
B(x,q)=e^{-2\pi \i \zeta_0-\pi \i \lambda x}
 \frac{\theta_1(\pi(x-\zeta_0+\frac{\tau}{2})\theta'_1(0)}
	{\theta_1(\pi(x+\frac{\tau}{2}))\theta_1(\pi\zeta_0)}
\sqrt{\frac{-\pi \theta'_1(0|\tau)}
{\frac{\d ^2 \text{log }\theta_1(\pi\zeta)}{\d \zeta ^2}
|_{\zeta=\zeta_0}}} \\
\tag 6.6\endgather$$

The factor $1-e^{4\pi\pi/\kappa}$ in front of the integral appears due
to the fact the cycle of integration passes through $\zeta_0$ twice.

Substituting this asymptotics in the equation (5.1), we get:

$$4\pi \i \frac{\frac{\d}{\d
\tau}A(q)}{A(q)} B(x,q) -2 \pi \i  \kappa \frac{\d}{\d\tau} B(x,q) +
\frac{\d^2}{\d x^2}B(x,q)=
2(\wp(x+\frac{\tau}{2}) + c) B(x,q),$$ or -- letting $\kappa\to 0$ --

$$\frac{\d^2}{\d x^2}B(x,q)= 2\left(\wp(x+\frac{\tau}{2}) + c -2\pi \i
\frac{\frac{\d}{\d\tau}A(q)}{A(q)}\right)B(x,q).\tag 6.7$$

This shows that the function $B(x,q)$, given by (6.6) is a solution of
Lam\'e equation with $m=1$. Therefore, we have proved the following
\proclaim{Proposition 6.1} Under the assumptions 1 and 2 above,
the function
$$f(x)=e^{-\pi \i \lambda x}\frac{\theta_1(\pi(x-\zeta_0+\frac{\tau}{2}))}
{\theta_1(\pi(x+\frac{\tau}{2}))},$$
where $\lambda$ and $\zeta_0$ are related by (6.4), is a solution of
Lam\'e equation of degree $m=1$. \endproclaim

Let us return to these assumptions. Consider first the case when
$q\in\R$, $\zeta_0=1/2$. One can easily check that in this case
$\lambda=0$, and that the function $|\theta_1(\pi\zeta|\tau)|$ reaches
at $\zeta=1/2$ its absolute maximum on the real line. Therefore, in
this case the assumptions 1 and 2 are satisfied. Therefore, in this
case the function $f(x)$ is indeed an asymptotics for the function
$F(x,q)$ and therefore is a solution of Lam\'e equation.

In general these assumptions are not true. However, the result is
still valid: the above defined function $f(x)$ is a solution of Lam\'e
equation. To see it, note that the assumprions 1,2
are satisfied for $q,\zeta_0$ in some neighborhood
of $q_0\in\R,\zeta_0=1/2$. Therefore, for all these $q,\zeta_0$ the
function $f(x)$ defined above is a solution of Lam\'e equation.  Since
$f$ analytically depends on $q,\zeta_0$ it implies that in fact $f(x)$
is a solution of Lam\'e equation for all $q,\zeta_0$ for which it
is well-defined (though in general we cannot claim that it is obtained
as an asymptotics of the correlation function $F(x,q)$).  So we have
obtained the following result (which is of course well-know, see \cite{WW}):
\proclaim{Proposition  6.2} For any $\lambda, \zeta_0$ satisfying (6.4),
the function
$$f(x)=e^{-\pi \i \lambda x}
\frac{\theta_1(\pi(x-\zeta_0+\frac{\tau}{2}))}
{\theta_1(\pi(x+\frac{\tau}{2}))},\tag 6.8$$
is a solution of Lam\'e equation of degree $m=1$. \endproclaim

It is an
instructive  exercise to check this by direct computation.

Let us consider some examples. If $\lambda=0$, it is easy to see that
the critical point is $\zeta_0=1/2$, and
$f(x)=G(1,-1|\tau,x+\frac{\tau}{2})=\sum\limits_{n\in\Z}
\frac{q^ne^{2\pi \i  nx}}{1+q^{2n}}=\frac{\bold K}{\pi}\text{dn }(2\bold
K x)$, where the periods $\bold K, \bold K'$ of Jacobi elliptic
functions are related with $q$ by $q=e^{-\pi \frac{\bold K'}{\bold
K}}$. More generally, if $\lambda\in 2\Z$ then the critical point is
$\zeta_0=\frac{1}{2}+\frac{\lambda}{2}\tau$, and the corresponding
function $f(x)$ is again $\text{dn }(2\bold K x)$.

Similarly, for $\lambda = 1$ the critical points are
$\zeta_0=\frac{\tau}{2},
\zeta_0=\frac{1}{2}+\frac{\tau}{2}$, and the function $f(x)=\text{sn }(2\bold
Kx), f(x)=\text{cn }(2\bold Kx)$; the same functions are obtained for
every odd integer $\lambda$. Note that we have two different solutions
corresponding to the same value of $\lambda$; in the best case only one of
them is an asymptotics of the correlation function on the torus; so the
interpretation of these solutions in terms of representation theory is
still unclear.

We have shown that $\lambda\in\Z$ gives us the classical periodic
solutions of Lam\'e equation of level $m=1$: $\text{dn }(2\bold
Kx),\text{sn }(2\bold Kx),\text{cn }(2\bold Kx)$ (which are of course
well known). The general theory (see \cite{WW}) says that there are no
other periodic solutions.Therefore, the asymptotics of the integral
representation of the function $F$ for $m=1$ gave us all the periodic
solutions of Lam\'e equation of degree 1.

The same technique can be applied for arbitrary $m$.
Though it is not easy to see what kind of contour we must take and
check the necessary conditions for the apllicability of the
multi-dimensional saddle point method, Proposition 6.2 suggests
the following formula:

\proclaim{Proposition 6.3} For any $m\in\Z_+$,
$\zeta=\zeta_1\ldots\zeta_m\in\C ^m, \lambda\in\C$ satisfying
$$\text{i}\lambda +m\frac{\theta'_1(\pi\zeta_i)}{\theta_1(\pi\zeta_i)}-
\sum\limits_{j\ne
i}\frac{\theta'_1(\pi(\zeta_i-\zeta_j))}{\theta_1(\pi(\zeta_i-\zeta_j))}=0
\text{ for all }i=1\ldots m,\tag 6.10$$

the function $f(x)$, defined by

$$f(x)=e^{-\pi \i  \lambda x}\prod\limits_{j=1}^m
\frac{\theta_1(\pi(x-\zeta_j+\frac{\tau}{2}))}
{\theta_1(\pi(x+\frac{\tau}{2}))} \tag 6.11$$

is a solution of the Lam\'e equation of degree $m$.
\endproclaim

This proposition can be proved by a direct calculation,
which shows that a
function of the form (6.11) is a solution of Lam\'e equation if and
only if $\lambda$ and $\zeta_i$ are related by a system of equations
(6.10) (see \cite{WW, \S 23.71}). It is also known that almost any solution
-- not necessarily periodic --
can be written in this form. The only exception is the case when the
constant $c$ in the Lam\'e equation (5.2) is an eigenvalue, i.e. when
this equation has a doubly-periodic solution. In this case the other
solution  is not periodic and cannot be expressed in terms of theta
functions. Therefore we see that  our approach gave us almost all
solutions of Lam\'e equation.

 This general formula has been known for a long
time (in \cite{WW} it is given with a reference to Hermite's note of
1872), though, of course, it was never obtained as an asymptotics of an
integral in the way we do.

\heading{\bf Conclusion} \endheading

We've considered in detail the case $\g=\sltwo$.  As for the
case of arbitrary $\g$, it is known there always exists an analogue of
Wakimoto realization (see \cite{FF}) and therefore, integral formulas
for intertwiners similar to (5.8) (see \cite{ATY}). This in turn
implies the existence of integral formulas for the traces $\Cal F$ of the
form (2.6). Therefore, it is natural to suggest the following
conjecture (for the case $n=1$, i.e. trace with insertion of one intertwiner):

\proclaim{Conjecture} If $V$ is a finite-dimensional representation
then the correlation function $\Cal F$ defined by (2.6) has the
following asymptotics as $\kappa\to 0$ (with all the derivatives):

$$\Cal F \sim C_1 \kappa^{N/2}C_2^{1/\kappa} \phi(x),$$
where $C_1,C_2$ do not depend on $x,\kappa$ (but may depend on $q,\lambda$).
\endproclaim

This would immediately imply that the function $\phi/\Cal F_0$
satisfies the equation

$$\Delta_{\frak h}\frac{\phi}{\Cal F_0}=
-\frac{1}{2\pi^2}
\sum\limits_{\alpha\in\Delta^+}\left(\wp\left(\frac{\tau|\alpha|}{g}
-\frac{h(\alpha)}{2\pi \i}\right)+c\right)f_\alpha e_\alpha
\frac{\phi}{\Cal F_0} ,\tag 6.12$$
which is obtained from (6.12) by letting $\kappa\to 0$.
This equation is a natural multidimensional analogue of the Lam\'e equation.
Moreover, if the highest weight $\lambda$ of the Verma module is
integral then the solutions we get must possess some remarkable
periodicity properties and should be considered as natural analogues
of Lam\'e functions. Note that in general $V_{\mu}(0)$ is not
one-dimensional, so $\phi$ is a vector function; however, for
$\g=\frak s \frak l _n$ there exist finite-dimensional modules with
one-dimensional  zero-weight component. In this case we obtain the
elliptic analogue of Calogero-Sutherland operator for the root system
$A_n$ (cf. \cite{OP}). This
will be discussed in detail in our future paper.

It is an interesting problem to consider the correlation functions in
the case where $V$ is not a finite-dimensional representation. For
example, for $\g=\sltwo$ one can define the module
$W_{\mu}=\{(x_1x_2)^\mu f(x_1,x_2)|f\text{ is a Laurent polynomial in
} x_1,x_2 \text{ of total degree zero}\}$ with
the action of $\sltwo$ given by
$$e=x_1\frac{\d}{\d x_2},\ \
h=x_1\frac{\d}{\d x_1}-x_2\frac{\d}{\d x_2}\ \
f=x_2\frac{\d}{\d x_1}.$$
In this form this module has an obvious generalization to the case
$\g=\frak s \frak l_n$.

It is easy to see that $W_\mu (0)$ is one-dimensional and
$ef|_{W_\mu(0)}=\mu(\mu+1)$. Also note that if $\mu\in\Z_+$ then this
module has a finite dimensional submodule of dimension $2\mu+1$.

Therefore, the traces of the form (2.6) for the intertwiners
$\Phi(z)\colon M_{\lambda,k}\to M_{\lambda,k}\otimes
W_\mu$  satisfy the parabolic PDE (5.14) with the constant $m=\mu$ --
an arbitrary complex number.
It is an interesting open question if these correlation functions have
asymptotics of the form

$$\Cal F \sim f(q,\lambda,\kappa )\phi(q,\lambda,x).$$
Then the function $\phi(q,\lambda,x)$ would be a
solution of the Lam\'e equation. This would provide a method of
constructing Lam\'e functions with
arbitrary complex value of the constant $m$ from representation theory
of $\widehat{\sltwo}$


\Refs

\ref\by[ATY] Awata, H., Tsuchia, A. and Yamada, Y. \paper Integral
formulas for the WZNW correlation functions \jour Nucl. Phys. \vol
B365 \yr 1991 \pages 680-696\endref

\ref\by[Ber] Bernard, D.\paper On the Wess-Zumino-Witten models on the
torus\jour Nucl. Phys.\vol B303\pages 77--93\yr 1988\endref

\ref\by[BF] Bernard, D., Felder, G.\paper Fock representations and
BRST cohomology in SL(2) current algebra\jour Commun. Math. Phys
\vol 127\pages 145-168\yr 1990\endref

\ref\by[CFW] Crivelli, M., Felder, G., Wieczerkowski, C.\paper
Generalized hypergeometric functions on the torus and adjoint
representation of $U_q({\frak s\frak l}_2)$, preprint\yr 1992\endref

\ref\by[Ch] Cherednik, I.V. \paper Integral solutions of
trigonometric Knizhnik-Zamolodchikov equations and Kac-Moody
algebras\jour Publ. RIMS, Kyoto University\vol 27(5)\pages727-744\yr
1991\endref

\ref\by[E] Etingof, P.I. \paper Representations of affine Lie algebras,
elliptic $r$-matrix systems, and special functions\jour hep-th
9303018; to appear in Commun. Math. Phys.\endref

\ref\by[FF] Feigin, B.L. and Frenkel, E.V. \jour Commun. Math. Phys.
\vol 128 \page 161 \yr 1990\endref

\ref\by[FR] Frenkel, I.B., and Reshetikhin, N.Yu.\paper Quantum affine
algebras and holonomic difference equations\jour Comm. Math. Phys.\vol
146\pages 1-60\yr 1992\endref


\ref\by[Koh] Kohno, T.\paper Monodromy representations of braid groups and
Yang-Baxter equations\jour Ann. Inst. Fourier\vol 37\pages 139-160
\yr 1987\endref

\ref\by[Kri] Krichever, I.M.\paper Methods of algebraic geometry in
the theory of non-linear equations\jour Russian Math. Surv. \vol
32:6\pages 185-213\yr 1977\endref

\ref\by[KZ] Knizhnik, V.G., and Zamolodchikov, A.B.\paper Current
algebra and Wess-Zumino model in two dimensions\jour Nucl. Phys.\vol
B247\pages 83-103\yr 1984\endref

\ref\by[OP] Olshanetsky, M.A.  and Perelomov, A.M.\paper Quantum
integrable systems related to Lie algebras \jour Phys. Rep. \vol 94
\pages 313-404\yr 1983
\endref

\ref\by[PS] Pressley, A., and Segal, G. \book Loop groups\publ
Clarendon Press\publaddr Oxford\yr 1986\endref

\ref\by[RV] Reshetikhin, N.Yu.  and Varchenko, A.N.\paper in preparation
\endref

\ref\by [SV] Schechtman, V.V., and Varchenko, A.N. \paper Arrangements of
hyperplanes and Lie algebra homology\jour Inv. Math.\vol 106\pages
134-194\yr 1991\endref

\ref\by[TK] Tsuchiya, A., Kanie, Y.\paper Vertex operators in
conformal field theory on $P^1$ and monodromy representations of braid
group\jour Adv. Stud. Pure Math.\vol 16\pages 297-372\yr 1988\endref

\ref\by[WW] Whittaker, E.T., Watson, G.N.\book Course of modern
analysis, 4th edition\publ Cambridge Univ. Press\yr 1958\endref
\endRefs
\enddocument
\end